\documentclass[onecolumn,showpacs,preprintnumbers,amsmath,amssymb]{revtex4}

\usepackage{amsmath}    
\usepackage{graphicx}   
\usepackage{graphics}
\usepackage{dcolumn}
\usepackage{bm}

\usepackage{latexsym}
\usepackage{amssymb}
\usepackage{amsmath}

\begin{document}
\title{Frustrated magnetic helices in MnSi-type crystals}
\author{Viacheslav A. Chizhikov\footnote{email: chizhikov@crys.ras.ru}, Vladimir E. Dmitrienko\footnote{email: dmitrien@crys.ras.ru}}
\affiliation{A.V.~Shubnikov Institute of Crystallography, 119333 Moscow, Russia}


\begin{abstract}

The spiral magnetic order in cubic MnSi-type
crystals is considered using the model of
classical Heisenberg ferromagnetics with an extra
interaction of the Dzyaloshinskii--Moriya (DM)
type between neighboring atoms. For all shortest
Mn-Mn bonds the DM vectors $\mathbf D$ are
expressed via the $\mathbf D$-vector for an
arbitrary chosen bond, in accordance with the
$P2_13$ space group of these crystals and bond
directions. The $\mathbf D$-vectors have the same
length and 24 different directions so that the
components of the $\mathbf D$-vectors are
changing similarly to the components of the
corresponding bond vectors. After minimization of
the magnetic energy it is found how the wave
vector $\mathbf{k}$ of magnetic helices depends
on $\mathbf D$: for any helix direction
$k=2(D_x-2D_y-D_z)/(3J)$ where $J$ is the
exchange interaction constant and $k$-vector is
in the units of the lattice constant. The wave
number $k$ determines both the sign and strength
of {\it global} spiraling whereas {\it locally},
within a unit cell, the helical order can be
strongly frustrated so that the twist angles
between neighboring ferromagnetic layers may be
even of different signs. Conical deformations of
helices caused by an arbitrary directed external
magnetic field is also considered within the same
model. The critical field of helix unwinding is
found and it is shown that even in the unwound
state there remains a residual periodic splay of
magnetic moments with the splay angle
proportional to $(D_x+D_z)/J$ which can be
measured by diffraction methods. The third
component of the $\mathbf D$-vector,
$D_x+D_y-D_z$, appears only in higher
approximations. It is also demonstrated how the
usually used continuous picture of moment
distribution can be obtained from the discrete
one in a coarse grain approximation.
\end{abstract}
\maketitle


\section{Introduction}
\label{sec:intro}

The spiral arrangements appear in very different
systems of condense matter physics and biology
\cite{Kornyshev} providing an important example
of a primitive self-organization. Sometimes this
self-organization results in rather intricate
structures (may be the first staircases to more
complicated biological patterns) like the blue
phases in chiral liquid crystals
\cite{Belyakov85,Wright89} or the Skyrmion
textures in chiral magnetics with $B20$ crystal
structure (MnSi, MnGe, Fe$_{1-x}$Co$_{x}$Si)
\cite{Bogdanov89,Rossler06,Yu11}. There is a
remarkable similarity in the physics of those so
different systems \cite{Tewari06}. Therefore it
is interesting and challenging to look for local
interactions responsible for macroscopic spiral
ordering and, in particular, to find
relationships between parameters of the local
interaction and the final structures. For
instance, the first question is what determines
the period and handedness of helices?

Most frequently, the helical arrangements are
described phenomenologically, just by adding
chiral terms like $\cal D \mathbf{n}\cdot {\rm
curl}\phantom{;}\mathbf{n}$ to the free energy of
the system where vector $\mathbf{n(r)}$ describes
a local system anisotropy. The period and
handedness of helices are determined by the
absolute value and sign of the pseudoscalar
coefficient $\cal D$; of course this term should
be allowed by the system symmetry (lack of
inversion is required). In the chiral liquid
crystals, the value of $\cal D$ is believed to be
determined by steric interactions between
neighboring chiral molecules but any quantitative
theory is still absent because of baffling
complexity of the problem. In cubic magnetic
crystals of the MnSi type, the helical ordering
is induced by the Dzyaloshinskii--Moriya (DM)
interaction between neighboring magnetic moments
$\mathbf{s}_i$ and $\mathbf{s}_j$
\cite{Nakanishi80,Bak80}. This interaction
\cite{Dzyaloshinskii57,Dzyaloshinskii58,Moriya60a,Moriya60b}
can exist both in centrosymmetric and
non-centrosymmetric crystals but only in the
latter case it results in long-range helical
structures. The most simple form of the DM term
in the interaction energy is written as
$\mathbf{D}_{ij} \cdot [ \mathbf{s}_i \times
\mathbf{s}_j ]$ where pseudovector coupling
coefficients $\mathbf{D}_{ij}$ originate from
superexchange interaction via neighboring atoms.

It was shown by Moriya that vectors
$\mathbf{D}_{ij}$ are determined by asymmetry of
local atomic arrangements. Recently it was
carefully investigated, using synchrotron
radiation and polarized neutron small-angle
diffraction, what is the relationship between the
handedness of the MnSi crystal structure and the
sign of magnetic helix
\cite{Grigoriev09,Grigoriev10}. It was
unequivocally demonstrated that the structural
chirality rigorously determines the magnetic
chirality of these compounds. The interplay
between structural chirality is also important
for spin chirality in multilayer structures
\cite{Bogdanov01,Grigoriev10b}. However, no
analytical formula related $\cal D$ and
$\mathbf{D}_{ij}$ has been found yet even for the
simple case of interaction between magnetic
moments of neighboring atoms. This problem is
indeed even more complicated because the
itinerant nature of magnetism in MnSi crystals.
Detailed numerical studies of the DM interaction
in MnSi-type crystals are performed by Hopkinson
and Kee \cite{Hopkinson} and their results will
be discussed below in comparison with the results
of the present paper.

A generic feature, intrinsic to complicated
spacial structures, is {\it frustration} of their
local ordering: the configurations with the best
local energy cannot be spread for all the system.
In the chiral liquid crystals and MnSi
heliomagnetics, the most favorable double-twist
local arrangements are restricted by topological
constrains existing in two-dimensional and
three-dimensional spaces
\cite{Belyakov85,Wright89}. Nevertheless, the
stable three-dimensional blue phases are possible
in a narrow range of temperatures whereas the
two-dimensional system of spin vertices can be
stabilized by external magnetic field. However,
in complicated crystals like La$_2$CuO$_4$, there
is an additional mechanism of frustrations
\cite{Shekhtman,Yildirim} related with different
orientations of $\mathbf{D}_{ij}$ for different
inter-spin bonds. We have found that this type of
frustration is also intrinsic to the MnSi
structure: at the atomic scale, spin rotation in
the magnetic helices can be strongly nonmonotonic
because of different orientations of
$\mathbf{D}_{ij}$ for different bonds. As a
result, unequal righthand and lefthand twists
usually coexist in each of the helices. For
instance, the coefficient $\cal D$ can become
zero even for nonzero $\mathbf{D}_{ij}$.

In this paper we first find a simple analytical
expression related the helix period and
handedness with the DM vectors of the shortest
Mn-Mn bonds. In this consideration, we adopt a
usually used approximation that the exchange
interaction is stronger than DM interaction and
twist angles between neighboring atoms are
correspondingly small. It is demonstrated that
for general orientation of DM vectors the MnSi
helical structure is frustrated so that the twist
angles between neighboring ferromagnetic layers
have different signs. Then, using the same
approximations, conical distortions of helices in
external magnetic fields are analyzed and the
critical field of helix unwinding is determined.
Finally, the parameters of the known
phenomenological description of MnSi helices are
related to the DM vectors of the microscopic
theory.

\section{Magnetic energy and crystal structure}
\label{sec:energy}

In a simple model of classical Heisenberg
ferromagnetics with an extra interaction of the
Dzyaloshinskii--Moriya (DM) type, the energy of
the system is expressed as a sum of pair
interactions between magnetic atoms and the
interaction of individual atoms with an external
magnetic field $\mathbf{H}$:
\begin{equation}
\label{eq:energy1}
E = \frac{1}{2} \sum_{i \neq j} ( -J_{ij} \mathbf{s}_i \cdot \mathbf{s}_j + \mathbf{D}_{ij} \cdot [ \mathbf{s}_i \times \mathbf{s}_j ] ) - g\mu_B \sum_i \mathbf{H} \cdot \mathbf{s}_i ,
\end{equation}
where the first summation is taken over all pairs
$\{ij\}$ of magnetic atoms, $J_{ij}$ is the
exchange interaction constant, $\mathbf{D}_{ij}$
are the Dzyaloshinskii--Moriya vectors, and
$g\mu_B$ is an effective magnetic moment.
Hereafter we will use the approximation of
classical spins with ferromagnetic exchange
coupling between nearest neighbors, $J_{ij}
\equiv J > 0$, and suppose that
$|\mathbf{s}_i|=1$ for all magnetic atoms. Using
the nearest neighbor approximation for the
MnSi-type crystals we can rewrite Eq.
(\ref{eq:energy1}) as
\begin{equation}
\label{eq:energy2}
E = \frac{1}{2} \sum_i \sum_{j=1}^6 ( -J \mathbf{s}_i \cdot \mathbf{s}_j + \mathbf{D}_{ij} \cdot [ \mathbf{s}_i \times \mathbf{s}_j ] ) - g\mu_B \sum_i \mathbf{H} \cdot \mathbf{s}_i ,
\end{equation}
where the first sum ($i$) is taken over all Mn
atoms of the crystal and the second sum ($j$) is
over six nearest Mn neighbors of $i$-th atom,
i.e. over six bonds connecting each Mn atom with
its nearest neighbors.

It should be emphasized that vectors
$\mathbf{D}_{ij}$ have different orientations for
different bonds and relations between different
$\mathbf{D}_{ij}$ can be found from symmetry
reasons. In the unit cell of MnSi-type crystals,
the magnetic atoms are at the $4a$ positions with
coordinates $(x,x,x)$, $(1-x, \frac{1}{2}+x,
\frac{1}{2}-x)$, $(\frac{1}{2}-x, 1-x,
\frac{1}{2}+x)$, $(\frac{1}{2}+x, \frac{1}{2}-x,
1-x)$ \cite{tables}. There are 24
crystallographically equivalent bonds between
nearest neighbors, and DM vectors for all bonds
can be obtained from DM vector of any bond using
symmetry transformations of the point group $23$
of MnSi crystal augmented with the inversion; the
latter appears owing to the permutation of atoms:
$\mathbf{D}_{ji} = -\mathbf{D}_{ij}$. The
procedure is, in principle, very simple: we
choose an arbitrary vector $(D_x,D_y,D_z)$ as DM
vector of the ``first'' bond, say $(-2x,
\frac{1}{2}, \frac{1}{2}-2x)$, connecting Mn atom
at $(x,x,x)$ with Mn atom at $(-x, \frac{1}{2}+x,
\frac{1}{2}-x)$. It should be an arbitrary vector
because no symmetry operation is associated with
the bond. Then, for this cubic space group, any
shortest bond between Mn atoms can be obtained
from the ``first'' one using cyclic permutations
(i.e. threefold rotations) and/or sign inversions
(i.e. twofold rotations and inversions) of its
vector components. The components of the
corresponding DM vector are expressed through
$D_x$, $D_y$ and $D_z$ by exactly the same cyclic
permutations and/or sign inversions. The list of
all the bonds and their $\mathbf{D}$-vectors is
given in Table \ref{tableD}. All
$\mathbf{D}_{ij}$ have the same length:
$|\mathbf{D}_{ij}|=D$.

It was shown \cite{Shekhtman,Yildirim} that the
DM term should be accompanied by two extra terms:
\begin{equation}
\label{eq:energyS} E^\prime = \frac{1}{2} \sum_{i
\neq j}
\left(\frac{\mathbf{D}^2_{ij}(\mathbf{s}_i\cdot\mathbf{s}_j)}{4J}-
\frac{(\mathbf{s}_i\cdot\mathbf{D}_{ij})(\mathbf{D}_{ij}\cdot
\mathbf{s}_j)}{2J}\right).
\end{equation}
These terms are of the same order of magnitude as
the DM term because for small DM interaction,
$|\mathbf{D}_{ij}| \ll J$, small angles between
$\mathbf{s}_i$ and $\mathbf{s}_j$ are of order
$|\mathbf{D}_{ij}|/J$. In Eq.
(\ref{eq:energyS}), with the same accuracy we can
put $\mathbf{s}_i=const$ for a small domain containing all
12 different interatomic bonds and averaging of
$(\mathbf{D}_{ij})_\alpha(\mathbf{D}_{ij})_\beta$
over these 12 bonds gives
$\frac{1}{3}D^2\delta_{\alpha\beta}$. Thus Eq.
(\ref{eq:energyS}) can be written as
\begin{equation}
\label{eq:energySS} E^\prime = \frac{1}{2}
\sum_{i \neq j}
\frac{D^2}{12J}\mathbf{s}_i\cdot\mathbf{s}_j,
\end{equation}
This term is isotropic and can be included
into exchange term of Eq. (\ref{eq:energy1}). We
see that in cubic crystals the extra terms given
by Eq. (\ref{eq:energyS}) are not important if
$|\mathbf{D}_{ij}| \ll J$.

Having a list of DM vectors one can calculate and
minimize numerically the magnetic energies
(\ref{eq:energy1}) and (\ref{eq:energy2}) for
rather big sets of atoms with distinct moment
configurations, including those similar to
Skyrmions. However in this paper we will use only
analytic calculations for detailed
consideration of helical structures.

\section{Helix structure in the absence of magnetic field}
\label{sec:helix}

Let us start with the simple case of free helices which provides a minimum of the magnetic energies (\ref{eq:energy1}) and (\ref{eq:energy2}) for $\mathbf{H}=0$. As it is known from numerous experiments, well below the Curie point the magnetic moments of Mn atoms form a simple helix. However, in the general case, the helices may be different for four different types of equivalent atomic positions in the unit cell (i.e. for positions with four different orientations of the local threefold axes), and this is confirmed by numerical simulations. Then, the spin structure can be described by the superposition of four helices
\begin{equation}
\label{eq:s1} \mathbf{s}^{t} =
\mathbf{n}^{t}_{cos} \cos\mathbf{k} \cdot
\mathbf{r} + \mathbf{n}^{t}_{sin} \sin\mathbf{k}
\cdot \mathbf{r} ,
\end{equation}
so that the condition $|\mathbf{s}^t|=1$ is
obviously obeyed. Here $\mathbf{r}$ is an atomic
position, $\mathbf{n}^{t}_{cos}$ and
$\mathbf{n}^{t}_{sin}$ are perpendicular unit
vectors, $\mathbf{k}$ is a wave vector. Upper
index $t = 1,2,3,4$ enumerates four atomic
positions in the ``first'' unit cell: $(x,x,x)$,
$(1-x, \frac{1}{2}+x, \frac{1}{2}-x)$,
$(\frac{1}{2}-x, 1-x, \frac{1}{2}+x)$,
$(\frac{1}{2}+x, \frac{1}{2}-x, 1-x)$,
correspondingly. We express all the atomic
positions in the units of the cubic unit cell so
that both $\mathbf{r}$ and $\mathbf{k}$ vectors
are dimensionless. Notice that all four spirals
given by Eq. (\ref{eq:s1}) have the same wave
vector whereas in the Skyrmion-like structures
there is a set of different wave vectors.

It can be easily seen that the spins of sublattice $t$ rotate in the plane perpendicular to vector
\begin{equation}
\label{eq:nrot}
\mathbf{n}^{t}_{rot} = [\mathbf{n}^{t}_{cos} \times \mathbf{n}^{t}_{sin}] ,
\end{equation}
whose direction generally speaking can differ from that of wave vector $\mathbf{k}$.

Now, using Eq. (\ref{eq:s1}) we can rewrite expression (\ref{eq:energy2}) for energy (at $\mathbf{H}=0$) as
\begin{equation}
\label{eq:Ehelix1}
\begin{array}{ll}
E_{helix} = & \frac{1}{2} \sum_i \sum_{j=1}^{6} \{ (-J \mathbf{n}_{cos}^{t(i)} \cdot \mathbf{n}_{cos}^{t(j)} + \mathbf{D}_{ij} \cdot [\mathbf{n}_{cos}^{t(i)} \times \mathbf{n}_{cos}^{t(j)}]) \cos\mathbf{k} \cdot \mathbf{r}_i \cos\mathbf{k} \cdot \mathbf{r}_j \\
 & + (-J \mathbf{n}_{cos}^{t(i)} \cdot \mathbf{n}_{sin}^{t(j)} + \mathbf{D}_{ij} \cdot [\mathbf{n}_{cos}^{t(i)} \times \mathbf{n}_{sin}^{t(j)}]) \cos\mathbf{k} \cdot \mathbf{r}_i \sin\mathbf{k} \cdot \mathbf{r}_j \\
 & + (-J \mathbf{n}_{sin}^{t(i)} \cdot \mathbf{n}_{cos}^{t(j)} + \mathbf{D}_{ij} \cdot [\mathbf{n}_{sin}^{t(i)} \times \mathbf{n}_{cos}^{t(j)}]) \sin\mathbf{k} \cdot \mathbf{r}_i \cos\mathbf{k} \cdot \mathbf{r}_j \\
 & + (-J \mathbf{n}_{sin}^{t(i)} \cdot \mathbf{n}_{sin}^{t(j)} + \mathbf{D}_{ij} \cdot [\mathbf{n}_{sin}^{t(i)} \times \mathbf{n}_{sin}^{t(j)}]) \sin\mathbf{k} \cdot \mathbf{r}_i \sin\mathbf{k} \cdot \mathbf{r}_j \} ,
\end{array}
\end{equation}
where upper index $t(i)$ designates ``type'' of
$i$-th atom or its belonging to one of four
sublattices, $t(i) = 1 \ldots 4$. The sines and
cosines products in Eq. (\ref{eq:Ehelix1}) can be
expressed through sines and cosines of arguments
$\mathbf{k} \cdot (2\mathbf{r}_i +
\mathbf{r}_{ij})$ and $\mathbf{k} \cdot
\mathbf{r}_{ij}$, with $\mathbf{r}_{ij}$ being
the distance between $i$-th and $j$-th atoms,
$\mathbf{r}_{ij} = \mathbf{r}_j - \mathbf{r}_i$.
It should be noted here that dependence on atomic
positions is present in these trigonometric
functions only. After summation over all unit
cells of the crystal all the functions of
argument $\mathbf{k} \cdot (2\mathbf{r}_i +
\mathbf{r}_{ij})$ vanish because $\mathbf{k}$
is not a reciprocal lattice vector, whereas the
functions of argument $\mathbf{k} \cdot
\mathbf{r}_{ij}$ do not depend on the unit cells
to which $i$-th and $j$-th atoms belong. Thus
we can rewrite expression (\ref{eq:Ehelix1}) for
energy as
\begin{equation}
\label{eq:Ehelix2}
\begin{array}{ll}
E_{helix} = & \frac{N_{cell}}{4} \sum_{i=1}^{4} \sum_{j=1}^{6} \{ (-J \mathbf{n}_{cos}^i \cdot \mathbf{n}_{cos}^{t(j)} + \mathbf{D}_{ij} \cdot [\mathbf{n}_{cos}^i \times \mathbf{n}_{cos}^{t(j)}]) \cos\mathbf{k} \cdot \mathbf{r}_{ij} \\
 & + (-J \mathbf{n}_{cos}^i \cdot \mathbf{n}_{sin}^{t(j)} + \mathbf{D}_{ij} \cdot [\mathbf{n}_{cos}^i \times \mathbf{n}_{sin}^{t(j)}]) \sin\mathbf{k} \cdot \mathbf{r}_{ij} \\
 & - (-J \mathbf{n}_{sin}^i \cdot \mathbf{n}_{cos}^{t(j)} + \mathbf{D}_{ij} \cdot [\mathbf{n}_{sin}^i \times \mathbf{n}_{cos}^{t(j)}]) \sin\mathbf{k} \cdot \mathbf{r}_{ij} \\
 & + (-J \mathbf{n}_{sin}^i \cdot \mathbf{n}_{sin}^{t(j)} + \mathbf{D}_{ij} \cdot [\mathbf{n}_{sin}^i \times \mathbf{n}_{sin}^{t(j)}]) \cos\mathbf{k} \cdot \mathbf{r}_{ij} \} ,
\end{array}
\end{equation}
where $N_{cell}$ is the number of unit cells in
the crystal, and first sum is now taken over four
magnetic atoms within one unit cell, so $t(i)=i$.

Eq. (\ref{eq:Ehelix2}) seems to be simple
enough for possible minimization. It depends on
15 variables only: three angles characterizing
orientation of triad $\{ \mathbf{n}_{cos}^t,
\mathbf{n}_{sin}^t, \mathbf{n}_{rot}^t \}$ for
each of four Mn sublattices, and three components
of the wave vector $\mathbf{k}$. The exchange
interaction constant $J$, DM vectors
$\mathbf{D}_{ij}$ and distances $\mathbf{r}_{ij}$
between neighboring atoms are supposed to be
invariable; we do not consider phenomena (e.g.
magnetostriction) changing these parameters.
Below, instead of numerical minimization, we will
look for approximate analytical solutions relying
on a small parameter.

Indeed, it is known from the experimental data
that the spin-orbit interaction is usually much
smaller than the exchange one, so that
$|\mathbf{D}_{ij}| \ll J$. In the MnSi crystal,
strong exchange interaction leads to the
ferromagnetic ordering, whereas small DM
interaction results in small canting of
neighboring magnetic moments. Small values of
$D/J$ and $|\mathbf{k}|$ (the latter is supposed
to be of the order of $D/J$) allows us to perform
approximate minimization of the energy. In
particular, we can take advantage of the fact,
that in this approximation the triads $\{
\mathbf{n}_{cos}^t, \mathbf{n}_{sin}^t,
\mathbf{n}_{rot}^t \}$ have close orientations
for four Mn sublattices.

Let us suppose that all triads $\{ \mathbf{n}_{cos}^t, \mathbf{n}_{sin}^t, \mathbf{n}_{rot}^t \}$ are close to triad $\{ \mathbf{n}_1, \mathbf{n}_2, \mathbf{n}_3 \}$, with $\mathbf{n}_3$ being a vector, which can differ from $\mathbf{n}_{\mathbf{k}} = \mathbf{k} / k$ ($k = \pm |\mathbf{k}|$ is a wave number; the sign defines the chirality of the helix), and $\{ \mathbf{n}_1, \mathbf{n}_2 \}$ being an orthonormal basis of the plane perpendicular to $\mathbf{n}_3$, $[\mathbf{n}_1 \times \mathbf{n}_2] = \mathbf{n}_3$. It is useful to introduce vector $\mathbf{n}_{3} \neq \mathbf{n}_{\mathbf{k}}$ for considering all partial helices on an equal footing and for possible future consideration of anisotropy effects.

The deviation of triad $\{ \mathbf{n}_{cos}^t, \mathbf{n}_{sin}^t, \mathbf{n}_{rot}^t \}$ from $\{ \mathbf{n}_1, \mathbf{n}_2, \mathbf{n}_3 \}$ can be considered as a sum of two small rotations: first one on the axis $\mathbf{n}_3$ by the angle $\omega_t$, and another one by the angle $\vec{\gamma}_t \perp \mathbf{n}_3$. Then, the vectors $\mathbf{n}_{cos}^t$, $\mathbf{n}_{sin}^t$, and $\mathbf{n}_{rot}^t$ can be represented as
\begin{equation}
\label{eq:basis2}
\left\{ \begin{array}{l}
\mathbf{n}_{cos}^t = (R^t \mathbf{n}_1) \cos\omega_t + (R^t \mathbf{n}_2) \sin\omega_t , \\
\mathbf{n}_{sin}^t = -(R^t \mathbf{n}_1) \sin\omega_t + (R^t \mathbf{n}_2) \cos\omega_t , \\
\mathbf{n}_{rot}^t = R^t \mathbf{n}_3 ,
\end{array} \right.
\end{equation}
with $R^t \equiv R(\vec{\gamma}_t)$ being the rotation matrix by the angle $\vec{\gamma}_t$. The substitution of Eq. (\ref{eq:basis2}) into Eq. (\ref{eq:Ehelix2}) gives
\begin{equation}
\label{eq:Ehelix3}
\begin{array}{ll}
E_{helix} = & \frac{N_{cell}}{4} \sum_{i=1}^{4} \sum_{j=1}^{6} \{ -J \{ (R^i \mathbf{n}_1) \cdot (R^{t(j)} \mathbf{n}_1) + (R^i \mathbf{n}_2) \cdot (R^{t(j)} \mathbf{n}_2) \} \cos(\omega_i - \omega_{t(j)} - \mathbf{k} \cdot \mathbf{r}_{ij}) \\
 & - J \{ (R^i \mathbf{n}_2) \cdot (R^{t(j)} \mathbf{n}_1) - (R^i \mathbf{n}_1) \cdot (R^{t(j)} \mathbf{n}_2) \} \sin(\omega_i - \omega_{t(j)} - \mathbf{k} \cdot \mathbf{r}_{ij}) \\
 & + \mathbf{D}_{ij} \cdot \{ [(R^i \mathbf{n}_1) \times (R^{t(j)} \mathbf{n}_1)] + [(R^i \mathbf{n}_2) \times (R^{t(j)} \mathbf{n}_2)] \} \cos(\omega_i - \omega_{t(j)} - \mathbf{k} \cdot \mathbf{r}_{ij}) \\
 & + \mathbf{D}_{ij} \cdot \{ [(R^i \mathbf{n}_2) \times (R^{t(j)} \mathbf{n}_1)] - [(R^i \mathbf{n}_1) \times (R^{t(j)} \mathbf{n}_2)] \} \sin(\omega_i - \omega_{t(j)} - \mathbf{k} \cdot \mathbf{r}_{ij}) \} .
\end{array}
\end{equation}
The seeming dependence on arbitrary vectors $\mathbf{n}_1$ and $\mathbf{n}_2$ can be eliminated with use of following two equations
\begin{subequations}     
\label{eq:useful1}
\begin{eqnarray}
(\mathbf{n}_1)_\alpha (\mathbf{n}_1)_\beta + (\mathbf{n}_2)_\alpha (\mathbf{n}_2)_\beta & = & \delta_{\alpha\beta} - (\mathbf{n}_3)_\alpha (\mathbf{n}_3)_\beta , \\
(\mathbf{n}_2)_\alpha (\mathbf{n}_1)_\beta - (\mathbf{n}_1)_\alpha (\mathbf{n}_2)_\beta & = & -\varepsilon_{\alpha\beta\gamma} (\mathbf{n}_3)_\gamma .
\end{eqnarray}
\end{subequations}
Then,
\begin{equation}
\label{eq:Ehelix4}
\begin{array}{ll}
E_{helix} = & \frac{N_{cell}}{4} \sum_{i=1}^{4} \sum_{j=1}^{6} \{ -J R^i_{\alpha\mu} R^{t(j)}_{\alpha\nu} (\delta_{\mu\nu} - (\mathbf{n}_3)_\mu (\mathbf{n}_3)_\nu) \cos(\omega_i - \omega_{t(j)} - \mathbf{k} \cdot \mathbf{r}_{ij}) \\
 & + J R^i_{\alpha\mu} R^{t(j)}_{\alpha\nu} \varepsilon_{\mu\nu\sigma} (\mathbf{n}_3)_\sigma \sin(\omega_i - \omega_{t(j)} - \mathbf{k} \cdot \mathbf{r}_{ij}) \\
 & + (\mathbf{D}_{ij})_\alpha \varepsilon_{\alpha\beta\lambda} R^i_{\beta\mu} R^{t(j)}_{\lambda\nu} (\delta_{\mu\nu} - (\mathbf{n}_3)_\mu (\mathbf{n}_3)_\nu) \cos(\omega_i - \omega_{t(j)} - \mathbf{k} \cdot \mathbf{r}_{ij}) \\
 & - (\mathbf{D}_{ij})_\alpha \varepsilon_{\alpha\beta\lambda} R^i_{\beta\mu} R^{t(j)}_{\lambda\nu} \varepsilon_{\mu\nu\sigma} (\mathbf{n}_3)_\sigma \sin(\omega_i - \omega_{t(j)} - \mathbf{k} \cdot \mathbf{r}_{ij}) \} .
\end{array}
\end{equation}
The phases $\omega$ are presented in expression (\ref{eq:Ehelix4}) only as differences, which is in agreement with the fact, that they are defined up to a constant term associated with the origin choice. The matrix $R(\vec{\gamma})$ can be expressed through the coordinates of the vector $\vec{\gamma}$:
\begin{equation}
\label{eq:useful2}
\begin{array}{ll}
(R(\vec{\gamma}))_{\alpha\beta} & = \cos\gamma \delta_{\alpha\beta} - \frac{\sin\gamma}{\gamma} \varepsilon_{\alpha\beta\sigma} \gamma_\sigma + \frac{1-\cos\gamma}{\gamma^2} \gamma_\alpha \gamma_\beta \\
 & \approx \delta_{\alpha\beta} - \varepsilon_{\alpha\beta\sigma} \gamma_\sigma + \frac{1}{2} (\gamma_\alpha \gamma_\beta - \delta_{\alpha\beta} \gamma^2) ,
\end{array}
\end{equation}
with $\gamma = |\vec{\gamma}|$. Then, we can rewrite the energy up to second order terms on $|\mathbf{k}| \sim (\omega_i - \omega_{t(j)}) \sim \gamma \sim (D/J)$:
\begin{equation}
\label{eq:Ehelix5}
E_{helix} = N_{cell} (-12J + \epsilon_1 + \epsilon_2) ,
\end{equation}
with
\begin{equation}
\label{eq:eps1-1}
\epsilon_1 = \frac{1}{2} \sum_{i=1}^{4} \sum_{j=1}^{6} \left\{ \frac{J}{2} (\omega_i - \omega_{t(j)} - \mathbf{k} \cdot \mathbf{r}_{ij})^2 - (\mathbf{D}_{ij} \cdot \mathbf{n}_3) (\omega_i - \omega_{t(j)} - \mathbf{k} \cdot \mathbf{r}_{ij}) \right\} ,
\end{equation}
\begin{equation}
\label{eq:eps2-1}
\epsilon_2 = \frac{1}{4} \sum_{i=1}^{4} \sum_{j=1}^{6} \left\{ \frac{J}{2} (\vec{\gamma}_i-\vec{\gamma}_{t(j)})^2 - \mathbf{D}_{ij} \cdot (\vec{\gamma}_i - \vec{\gamma}_{t(j)}) \right\} .
\end{equation}
As we can see, the energy breaks up into two contributions depending on $\omega$ and $\vec{\gamma}$ and related to each other only by direction of the axis $\mathbf{n}_3$.

The routine minimization of $\epsilon_1$ as a
function of independent variables
$(\omega_2-\omega_1)$, $(\omega_3-\omega_1)$ and
$(\omega_4-\omega_1)$ (performed with Wolfram
Research Mathematica) gives
\begin{equation}
\label{eq:omega1}
\left\{
\begin{array}{lll}
\omega_2 - \omega_1 & = & -\frac{1}{4} (k_x + k_z) (1-8x) - \frac{D_x+D_z}{2J} ((\mathbf{n}_3)_x + (\mathbf{n}_3)_z) , \\
\omega_3 - \omega_1 & = & -\frac{1}{4} (k_x + k_y) (1-8x) - \frac{D_x+D_z}{2J} ((\mathbf{n}_3)_x + (\mathbf{n}_3)_y) , \\
\omega_4 - \omega_1 & = & -\frac{1}{4} (k_y + k_z) (1-8x) - \frac{D_x+D_z}{2J} ((\mathbf{n}_3)_y + (\mathbf{n}_3)_z) ,
\end{array}
\right.
\end{equation}
and
\begin{equation}
\label{eq:eps1-2}
\epsilon_1 = J \left( -\frac{(D_x+D_z)^2}{J^2} + \frac{3}{4} k^2 - k \frac{D_x-2D_y-D_z}{J} (\mathbf{n}_{\mathbf{k}} \cdot \mathbf{n}_3) \right) .
\end{equation}
It is natural that $\epsilon_1$ does not depend on $x$, because real atomic coordinates are not involved in energy definition (\ref{eq:energy1}). From the physical point of view it is more interesting that $\epsilon_1$ is also independent of orientation of wave vector $\mathbf{k}$ relative to coordinate axes. This fact is in a good accordance with experimental evidence of the possibility of easy reorientation of helix axis in MnSi by magnetic field, surface anchoring or other effects.

The minimization of the expression (\ref{eq:eps1-2}) on $k$ gives us the following value for the wave number depending on DM vector $\mathbf{D}$ and rotation axis orientation relative to $\mathbf{k}$
\begin{equation}
\label{eq:k1}
k(\mathbf{D}, \widehat{\mathbf{n}_{\mathbf{k}}\mathbf{n}_3}) = \frac{2}{3} \left( \frac{D_x-2D_y-D_z}{J} \right) (\mathbf{n}_{\mathbf{k}} \cdot \mathbf{n}_3) ,
\end{equation}
and the part $\epsilon_1$ of unit cell magnetic energy can be finally expressed as
\begin{equation}
\label{eq:eps1-3}
\epsilon_1 = J \left( -\frac{(D_x+D_z)^2}{J^2} - \frac{1}{3} \frac{(D_x-2D_y-D_z)^2}{J^2} (\mathbf{n}_{\mathbf{k}} \cdot \mathbf{n}_3)^2 + \frac{3}{4} (k - k(\mathbf{D}, \widehat{\mathbf{n}_{\mathbf{k}}\mathbf{n}_3}))^2 \right) ,
\end{equation}
thereby the elasticity of the helix is $\frac{3}{4}J$.

It is evident from Eq. (\ref{eq:eps1-3}) that
$\epsilon_1$ has minimum, when $\mathbf{n}_3$
coincides with $\mathbf{n}_{\mathbf{k}}$, but it
should be noted that, when $\mathbf{n}_3$
deviates from $\mathbf{n}_{\mathbf{k}}$ by a
small angle $\gamma \sim (D/J)$, the energy
obtains only a small positive addition of the
order of $J(D/J)^4$. Thus, there is a soft mode
associated with deviation of rotation axis from
helix direction. This deviation gives a
cycloidal component to the helix.

Eq. (\ref{eq:eps2-1}) for $\epsilon_2$
contains vectors $\vec{\gamma}_t$ only within
differences, so, in analogy with $\omega$, the
vectors $\vec{\gamma}_t$ are defined up to a
constant vector perpendicular to $\mathbf{n}_3$.
However, unlike the case of $\omega$, where a
common term added to all phases results only in
redefining of the origin, the constant vector
added to all vectors $\vec{\gamma}_t$ turns
rotation axes for all four sublattices by the
same angle. In fact, this turning has been
already taken into account in the dependence of
$\epsilon_1$ on $(\mathbf{n}_{\mathbf{k}} \cdot
\mathbf{n}_3)$. We should also remind that Eq.
(\ref{eq:eps2-1}) has been obtained in the
assumption that $\gamma \sim (D/J)$. In order to
avoid the ambiguity of definition of vectors
$\vec{\gamma}_t$, we can use additional condition
$\sum_{t=1}^4 \vec{\gamma}_t = 0$.

The routine minimization of $\epsilon_2$ on independent variables $(\vec{\gamma}_2-\vec{\gamma}_1)$, $(\vec{\gamma}_3-\vec{\gamma}_1)$ and $(\vec{\gamma}_4-\vec{\gamma}_1)$ (performed with Wolfram Research Mathematica) gives
\begin{equation}
\label{eq:gamma1}
\left\{
\begin{array}{lll}
\vec{\gamma}_2 - \vec{\gamma}_1 & = & \frac{D_x+D_z}{2J} \{ (\mathbf{n}_3)_x (\mathbf{n}_3)_z - (\mathbf{n}_3)_y^2 - (\mathbf{n}_3)_z^2 , \\
& & (\mathbf{n}_3)_x (\mathbf{n}_3)_y + (\mathbf{n}_3)_y (\mathbf{n}_3)_z , (\mathbf{n}_3)_x (\mathbf{n}_3)_z - (\mathbf{n}_3)_x^2 - (\mathbf{n}_3)_y^2 \} , \\
\vec{\gamma}_3 - \vec{\gamma}_1 & = & \frac{D_x+D_z}{2J} \{ (\mathbf{n}_3)_x (\mathbf{n}_3)_y - (\mathbf{n}_3)_y^2 - (\mathbf{n}_3)_z^2 , \\
& & (\mathbf{n}_3)_x (\mathbf{n}_3)_y - (\mathbf{n}_3)_x^2 - (\mathbf{n}_3)_z^2 , (\mathbf{n}_3)_x (\mathbf{n}_3)_z + (\mathbf{n}_3)_y (\mathbf{n}_3)_z \} , \\
\vec{\gamma}_4 - \vec{\gamma}_1 & = & \frac{D_x+D_z}{2J} \{ (\mathbf{n}_3)_x (\mathbf{n}_3)_y + (\mathbf{n}_3)_x (\mathbf{n}_3)_z , \\
& & (\mathbf{n}_3)_y (\mathbf{n}_3)_z - (\mathbf{n}_3)_x^2 - (\mathbf{n}_3)_z^2 , (\mathbf{n}_3)_y (\mathbf{n}_3)_z - (\mathbf{n}_3)_x^2 - (\mathbf{n}_3)_y^2 \} ,
\end{array}
\right.
\end{equation}
and
\begin{equation}
\label{eq:eps2-2}
\epsilon_2 = -\frac{(D_x+D_z)^2}{J} .
\end{equation}
The full energy of the helix can be now written as

\begin{equation}
\label{eq:Ehelix6}
E_{helix} = N_{cell} J \left( -12 - 2\frac{(D_x+D_z)^2}{J^2} - \frac{1}{3} \frac{(D_x-2D_y-D_z)^2}{J^2} (\mathbf{n}_{\mathbf{k}} \cdot \mathbf{n}_3)^2 + \frac{3}{4} (k - k(\mathbf{D}, \widehat{\mathbf{n}_{\mathbf{k}}\mathbf{n}_3}))^2 \right) ,
\end{equation}
where the contribution $-12J$ per unit cell is associated with ferromagnetic ordering of spins (12 is the number of Mn-Mn bonds per unit cell), the term $-2\frac{(D_x+D_z)^2}{J}$, combining $\epsilon_2$ with a part of $\epsilon_1$, describes the energy of divergence between sublattices (the energy of canting), and remaining terms represent the twist energy of the helix.

The minimum of the energy (\ref{eq:Ehelix6}) is achieved, when $\mathbf{n}_3 = \mathbf{n}_\mathbf{k}$. Then,
\begin{equation}
\label{eq:Ehelix7}
E_{helix} = N_{cell} J \left( -12 - 2\frac{(D_x+D_z)^2}{J^2} - \frac{1}{3} \frac{(D_x-2D_y-D_z)^2}{J^2} + \frac{3}{4} (k - k(\mathbf{D}))^2 \right) ,
\end{equation}
where
\begin{equation}
\label{eq:k2}
k(\mathbf{D}) = \frac{2}{3} \left( \frac{D_x-2D_y-D_z}{J} \right) .
\end{equation}

It is also useful to write the full divergence vectors for sublattices using the formula
\begin{equation}
\label{eq:psi1}
\vec{\psi} = \omega\mathbf{n}_3 + \vec{\gamma} .
\end{equation}
then, from Eqs. (\ref{eq:omega1}) and (\ref{eq:gamma1}) follows
\begin{equation}
\label{eq:psi2}
\left\{
\begin{array}{lll}
\vec{\psi}_2 - \vec{\psi}_1 & = & \frac{\sqrt{3}}{4} \frac{D_x+D_z}{J} (\mathbf{e}_2 - \mathbf{e}_1) + \frac{\sqrt{3}}{8} (1-8x) (\mathbf{k} \cdot \mathbf{e}_2 - \mathbf{k} \cdot \mathbf{e}_1) \mathbf{n}_3 , \\
\vec{\psi}_3 - \vec{\psi}_1 & = & \frac{\sqrt{3}}{4} \frac{D_x+D_z}{J} (\mathbf{e}_3 - \mathbf{e}_1) + \frac{\sqrt{3}}{8} (1-8x) (\mathbf{k} \cdot \mathbf{e}_3 - \mathbf{k} \cdot \mathbf{e}_1) \mathbf{n}_3 , \\
\vec{\psi}_4 - \vec{\psi}_1 & = & \frac{\sqrt{3}}{4} \frac{D_x+D_z}{J} (\mathbf{e}_4 - \mathbf{e}_1) + \frac{\sqrt{3}}{8} (1-8x) (\mathbf{k} \cdot \mathbf{e}_4 - \mathbf{k} \cdot \mathbf{e}_1) \mathbf{n}_3 ,
\end{array}
\right.
\end{equation}
or, using the additional condition $\sum_{t=1}^4 \vec{\psi}_t = 0$,
\begin{equation}
\label{eq:psi3}
\vec{\psi}_t = \frac{\sqrt{3}}{4} \frac{D_x+D_z}{J} \mathbf{e}_t + \frac{\sqrt{3}}{8} (1-8x) (\mathbf{k} \cdot \mathbf{e}_t) \mathbf{n}_3 ,
\end{equation}
where $\mathbf{e}_1$, $\mathbf{e}_2$, $\mathbf{e}_3$ and $\mathbf{e}_4$ are the unit vectors directed along 3-fold symmetry axes corresponding to four sublattices:
\begin{equation}
\label{eq:e}
\left\{
\begin{array}{l}
\mathbf{e}_1 = \frac{1}{\sqrt{3}} (1, 1, 1) , \\
\mathbf{e}_2 = \frac{1}{\sqrt{3}} (-1, 1, -1) , \\
\mathbf{e}_3 = \frac{1}{\sqrt{3}} (-1, -1, 1) , \\
\mathbf{e}_4 = \frac{1}{\sqrt{3}} (1, -1, -1) .
\end{array}
\right.
\end{equation}

Finally, we can express vectors $\mathbf{n}_{cos}$, $\mathbf{n}_{sin}$ and $\mathbf{n}_{rot}$ for each of the sublattices:
\begin{equation}
\label{eq:3n}
\left\{
\begin{array}{l}
\mathbf{n}_{cos}^t = \mathbf{n}_1 + [\vec{\psi}_t \times \mathbf{n}_1] = \mathbf{n}_1 + \frac{\sqrt{3}}{4} \frac{D_x+D_z}{J} [\mathbf{e}_t \times \mathbf{n}_1] + \frac{\sqrt{3}}{8} (1-8x) (\mathbf{k} \cdot \mathbf{e}_t) \mathbf{n}_2 , \\
\mathbf{n}_{sin}^t = \mathbf{n}_2 + [\vec{\psi}_t \times \mathbf{n}_2] = \mathbf{n}_2 + \frac{\sqrt{3}}{4} \frac{D_x+D_z}{J} [\mathbf{e}_t \times \mathbf{n}_2] - \frac{\sqrt{3}}{8} (1-8x) (\mathbf{k} \cdot \mathbf{e}_t) \mathbf{n}_1 , \\
\mathbf{n}_{rot}^t = \mathbf{n}_3 + [\vec{\psi}_t \times \mathbf{n}_3] = \mathbf{n}_3 + \frac{\sqrt{3}}{4} \frac{D_x+D_z}{J} [\mathbf{e}_t \times \mathbf{n}_3] .
\end{array}
\right.
\end{equation}

\begin{figure}[h]
\includegraphics[width=8cm]{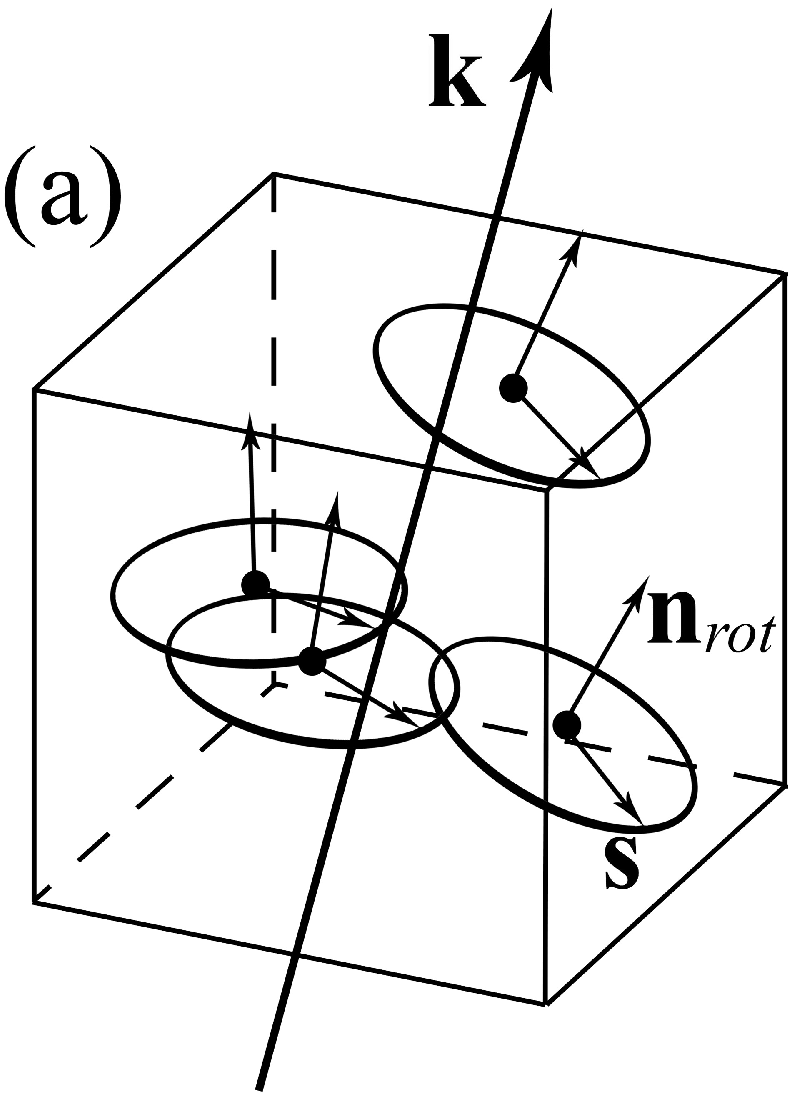}
\includegraphics[width=8cm]{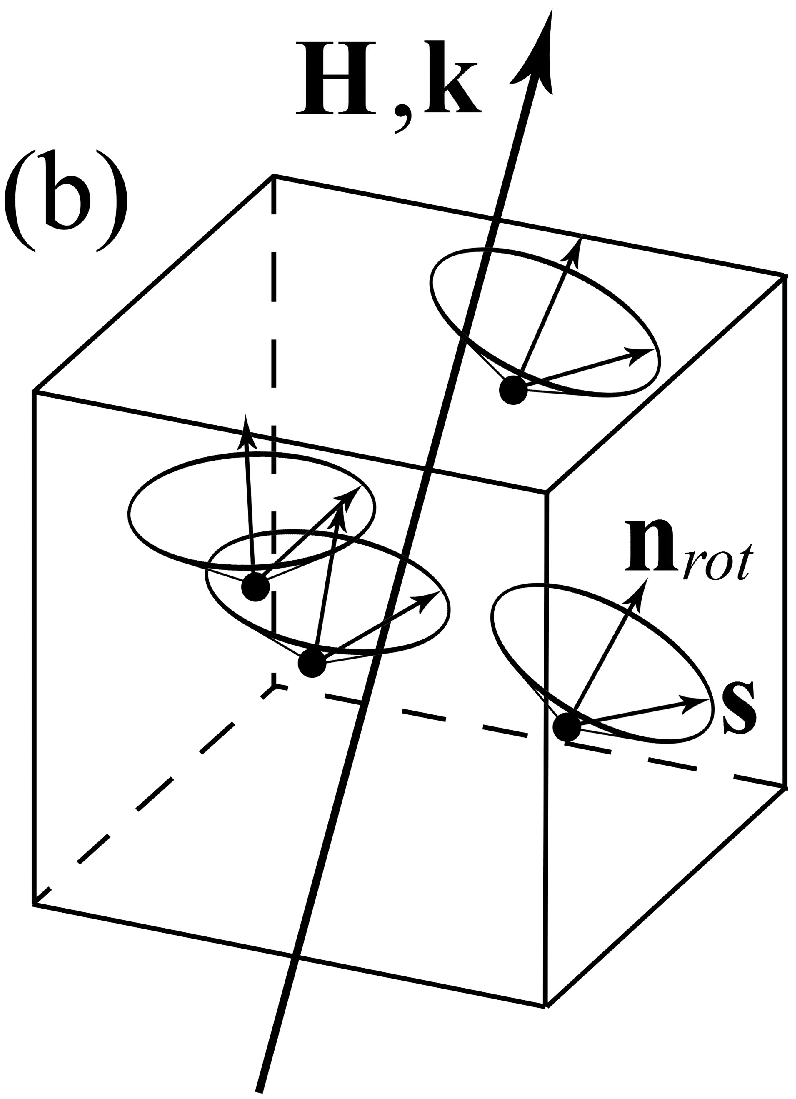}
\caption{\label{fig1}(a) Organization of a free
helix with an arbitrary wave vector $\mathbf{k}$
(no magnetic field). The spins of each of four
sublattices rotate around their own rotation
directions $\mathbf{n}_{rot}$ (see Eq.
(\ref{eq:3n})), which differ slightly from that
of $\mathbf{k}$. In the considered model, the
energy of the free helix does not depend on
direction of $\mathbf{k}$ whereas experimentally
the free helix is oriented either along $[111]$
or $[001]$ direction owing to the crystal field
anisotropy. (b) Conical helix spin structure in
an external magnetic field. The spins of each of
four sublattices have a component directed along
corresponding vector $\mathbf{n}_{rot}$; its
value is proportional to $|\mathbf{H}|/H_c$ for
$|\mathbf{H}| \le H_c$ and is equal to 1 for
$|\mathbf{H}| \ge H_c$; $\mathbf{k}$-vector is
directed along the magnetic field.}
\end{figure}

Thus, in the common case each sublattice has its own rotation axis different from $\mathbf{n}_\mathbf{k}$ as it shown in Fig.~\ref{fig1}(a). The divergence of the axes is defined by Eqs.~(\ref{eq:3n}). The substitution of Eqs.~(\ref{eq:3n}) into Eq.~(\ref{eq:s1}) gives the spin of an individual Mn atom
\begin{equation}
\label{eq:s3}
\begin{array}{ll}
\mathbf{s}_i = & \left( \mathbf{n}_1 + \frac{\sqrt{3}}{4} \frac{D_x+D_z}{J} [\mathbf{e}_{t(i)} \times \mathbf{n}_1] + \frac{\sqrt{3}}{8} (1-8x) (\mathbf{k} \cdot \mathbf{e}_{t(i)}) \mathbf{n}_2 \right) \cos \mathbf{k} \cdot \mathbf{r}_i \\
 & + \left( \mathbf{n}_2 + \frac{\sqrt{3}}{4} \frac{D_x+D_z}{J} [\mathbf{e}_{t(i)} \times \mathbf{n}_2] - \frac{\sqrt{3}}{8} (1-8x) (\mathbf{k} \cdot \mathbf{e}_{t(i)}) \mathbf{n}_1 \right) \sin \mathbf{k} \cdot \mathbf{r}_i ,
\end{array}
\end{equation}
The question can arise: why parameter $x$ appears in Eq.~(\ref{eq:s3}), in spite of its absence in the initial Eq.~(\ref{eq:energy2})? In fact, $x$ is brought in by Eq.~(\ref{eq:s1}) depending on real atomic positions. Indeed, parameter $x$ is implicitly involved with atomic coordinates $\mathbf{r}_i$, and expressing $\mathbf{r}$ as
\begin{equation}
\label{eq:r}
\mathbf{r} = \mathbf{p} + \mathbf{r}^\prime ,
\end{equation}
with $\mathbf{p}$ being a period of the crystal or the origin of a unit cell, and $\mathbf{r}^\prime \in \{ (x,x,x), (1-x, \frac{1}{2}+x, \frac{1}{2}-x), (\frac{1}{2}-x, 1-x, \frac{1}{2}+x), (\frac{1}{2}+x, \frac{1}{2}-x, 1-x) \}$ being an atomic position within the cell, and using the condition $\mathbf{k} \cdot \mathbf{r}^\prime \ll 1$, we can rewrite Eq.~(\ref{eq:s3}) as
\begin{equation}
\label{eq:s4}
\begin{array}{ll}
\mathbf{s}_i = & \left( \mathbf{n}_1 + \frac{\sqrt{3}}{4} \frac{D_x+D_z}{J} [\mathbf{e}_{t(i)} \times \mathbf{n}_1] + \left\{ \frac{\sqrt{3}}{8} (1-8x) (\mathbf{k} \cdot \mathbf{e}_{t(i)}) + \mathbf{k} \cdot \mathbf{r}^\prime_{t(i)} \right\} \mathbf{n}_2 \right) \cos \mathbf{k} \cdot \mathbf{p}_i \\
 & + \left( \mathbf{n}_2 + \frac{\sqrt{3}}{4} \frac{D_x+D_z}{J} [\mathbf{e}_{t(i)} \times \mathbf{n}_2] - \left\{ \frac{\sqrt{3}}{8} (1-8x) (\mathbf{k} \cdot \mathbf{e}_{t(i)}) + \mathbf{k} \cdot \mathbf{r}^\prime_{t(i)} \right\} \mathbf{n}_1 \right) \sin \mathbf{k} \cdot \mathbf{p}_i ,
\end{array}
\end{equation}
where parameter $x$ vanishes in curly brackets after substitution of corresponding $\mathbf{e}$ and $\mathbf{r}^\prime$.

\begin{figure}[h]
\includegraphics[width=8cm]{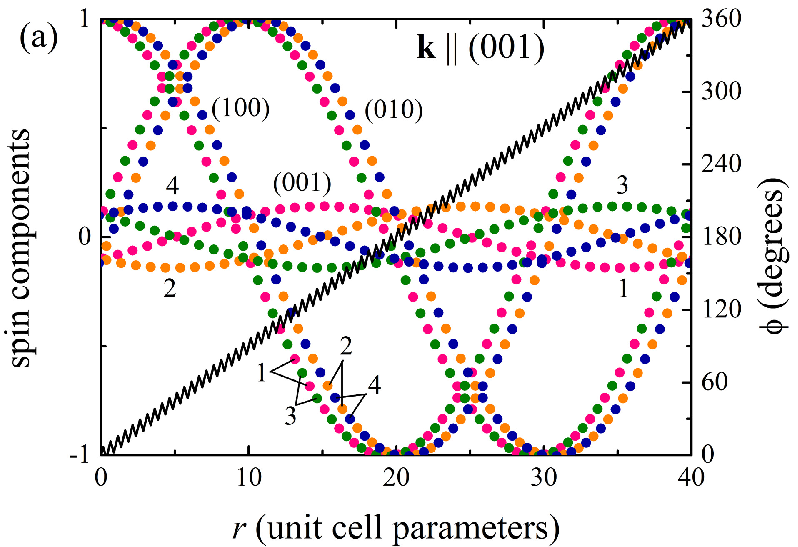}
\includegraphics[width=8cm]{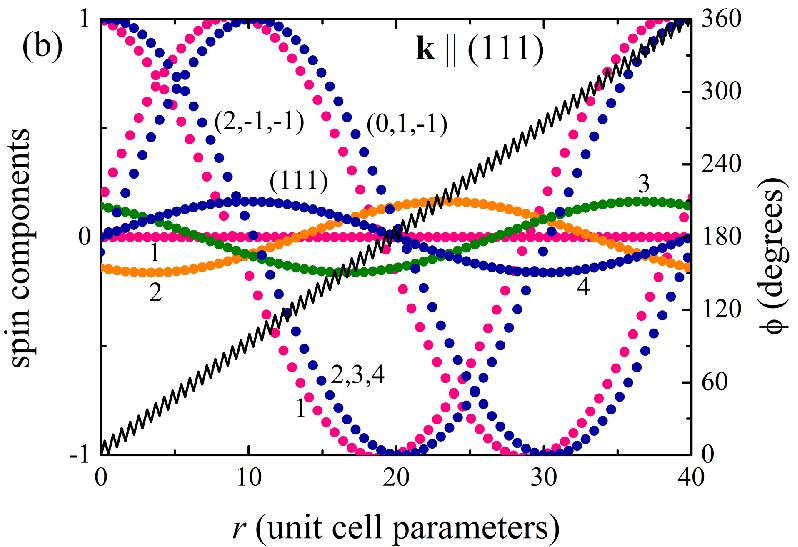}
\includegraphics[width=8cm]{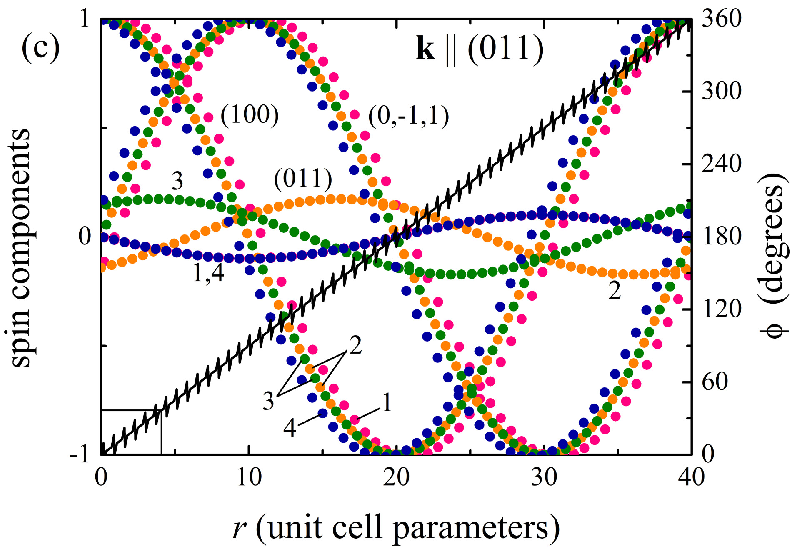}
\includegraphics[width=8cm]{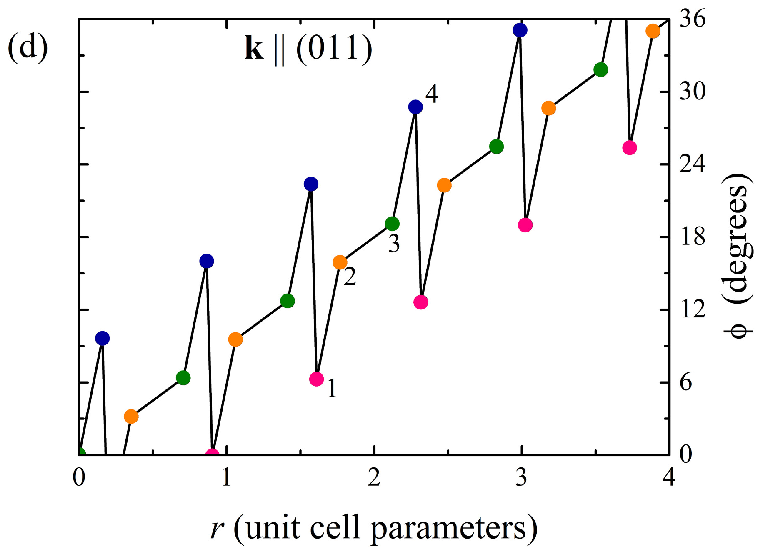}
\caption{(a)-(c) Details of spin helix structures calculated with the use of Eq. (\ref{eq:s3}) for the wave vectors $\mathbf{k}$ directed along $(001)$, $(111)$ and $(011)$ crystal axes, correspondingly. The wave number $|\mathbf{k}|$ is chosen to be equal to $2\pi /40$, which is close to the experimental value for MnSi crystal where the helix period is approximately 40 unit cells. The canting $\mathbf{D}$-vector component is taken to be rather large, $D_x+D_z = 0.4 J$, to emphasize nonmonotonic frustrated behavior of helices. Three spin components along directions $\mathbf{n}_1$, $\mathbf{n}_2$ and $\mathbf{n}_3 = \mathbf{n}_\mathbf{k}$ are presented by points with four colors corresponding to four sublattices. The black saw represents the angle $\phi$ between spins in the successive ferromagnetic planes, $\tan\phi = (\mathbf{s} \cdot \mathbf{n}_2) / (\mathbf{s} \cdot \mathbf{n}_1)$. (d) An enlarged part of the saw restricted by a small box in (c); the line is drown to guide the eyes.}  \label{fig2}
\end{figure}

Fig.~\ref{fig2} shows spin helix structures calculated with the use of Eq.~(\ref{eq:s3}) for directions $(001)$, $(111)$ and $(011)$ of spiral axis in the case of $\mathbf{n}_3 = \mathbf{n}_\mathbf{k}$. The wave number $|\mathbf{k}|$ is chosen to be equal to $2\pi /40$, which is close to the experimental value for MnSi crystal. The canting $\mathbf{D}$-vector component is taken to be rather large, $D_x+D_z = 0.4 J$, in order to emphasize nonmonotonic frustrated behavior of helices. In this case
\begin{equation}
\label{eq:fig2condition}
\frac{|D_x+D_z|}{4J} \gg \frac{(8x-1) |\mathbf{k}|}{8} ,
\end{equation}
and the main contribution in a canting is due to the terms with $[\mathbf{e}_{t(i)} \times \mathbf{n}_{1,2}]$.

The extended sinusoids with phase shift of $\frac{\pi}{2}$ represent spin projections on directions $\mathbf{n}_1$ and $\mathbf{n}_2$, the low ones do spin components along the wave vector, which appear due to the canting. The coloured circles correspond to magnetic atoms in four sublattices of the crystal. Being all different in the common case, the sinusoids connected with different sublattices can coincide for certain directions of $\mathbf{k}$ reflecting the symmetry. For instance, in the case of $\mathbf{k} \parallel (001)$ (Fig.~\ref{fig2}(a)) the extended sinusoids coincide, which correspond to sublattices 1, 3 and 2, 4; but the circles of different colors do not coincide because the atoms belonging to different sublattices lie in different planes perpendicular to $\mathbf{k}$. On the contrary, in the case of $\mathbf{k} \parallel (111)$ (Fig.~\ref{fig2}(b)) coincide both the extended sinusoids and related circles corresponding to sublattices 2, 3, 4, because their sites lie in the same atomic planes perpendicular to $\mathbf{k}$. Both two cases, (a) and (b), due to their symmetry ((001) is the axis $2_1$ and (111) is the axis $3$ of the crystal) are characterized by only two extended sinusoids, which defines the view of the black saws representing the angle $\phi$ between spins in the successive ferromagnetic planes, $\tan\phi = (\mathbf{s} \cdot \mathbf{n}_2) / (\mathbf{s} \cdot \mathbf{n}_1)$. The symmetry is also reflected in the low sinusoids corresponding to the spin components parallel to $\mathbf{k}$.

When $\mathbf{k} \parallel (011)$ (Fig.~\ref{fig2}(c)) there are three extended sinusoids, which is also reflected in the view of the cooresponding black saw (Fig.~\ref{fig2}(d)). In this case, coincide the extended sinusoids corresponding to sublattices 2, 3 and the low ones corresponding to sublattices 1, 4. It may seem from Fig.~\ref{fig2}(c) that sublattices 1 and 4 have the same projections on $\mathbf{k}$, but it is evident from Fig.~\ref{fig2}(d) that it is not so. Indeed, the atoms of types 1 and 4 lie in close but different planes; their proximity is due to the approximate equality $4x \approx \frac{1}{2}$.

The ``saws'' in the figures show a frustration which is the intrinsic feature of the helices due to the spin canting. Owing to the frustration the spins in successive ferromagnetic planes can rotate inversely to twist direction of the helix.

\section{Helix in magnetic field}
\label{sec:magnetic}

Let us return to the case of nonzero magnetic field $\mathbf{H}$. Numerous experimental data, as well as the results obtained with the use of phenomenological theory, show that in this case the magnetic moments of Mn atoms form a conical helix. The peculiarity of such structure is that the field of magnetic moment posesses a constant component directed along $\mathbf{H}$. Then, the spin structure can be described by the following expression
\begin{equation}
\label{eq:s2} \mathbf{s}^{t} = \mathbf{n}_{rot}^t
\sin\xi + \mathbf{n}_{cos}^t \cos\xi
\cos\mathbf{k} \cdot \mathbf{r} +
\mathbf{n}_{sin}^t \cos\xi \sin\mathbf{k} \cdot
\mathbf{r} ,
\end{equation}
which satisfies the condition $|\mathbf{s}^t|=1$. Here $\xi$ is an angle characterizing the ratio between constant and rotating components of the magnetic moments. The substitution of Eq. (\ref{eq:s2}) in the expression (\ref{eq:energy2}) and summation over all the unit cells of the crystal give
\begin{equation}
\label{eq:EhelixH1}
\begin{array}{ll}
E_{helix,H} = & \frac{N_{cell}}{2} \sum_{i=1}^{4} \sum_{j=1}^{6} ( -J \mathbf{n}_{rot}^{i} \cdot \mathbf{n}_{rot}^{t(j)} + \mathbf{D}_{ij} \cdot [\mathbf{n}_{rot}^{i} \times \mathbf{n}_{rot}^{t(j)}] ) \sin\xi_{i} \sin\xi_{t(j)} \\
 & + \frac{N_{cell}}{4} \sum_{i=1}^{4} \sum_{j=1}^{6} \{ ( -J \mathbf{n}_{cos}^{i} \cdot \mathbf{n}_{cos}^{t(j)} + \mathbf{D}_{ij} \cdot [\mathbf{n}_{cos}^{i} \times \mathbf{n}_{cos}^{t(j)}] ) \cos\mathbf{k} \cdot \mathbf{r}_{ij} \\
 & + ( -J \mathbf{n}_{cos}^{i} \cdot \mathbf{n}_{sin}^{t(j)} + \mathbf{D}_{ij} \cdot [\mathbf{n}_{cos}^{i} \times \mathbf{n}_{sin}^{t(j)}] ) \sin\mathbf{k} \cdot \mathbf{r}_{ij} \\
 & - ( -J \mathbf{n}_{sin}^{i} \cdot \mathbf{n}_{cos}^{t(j)} + \mathbf{D}_{ij} \cdot [\mathbf{n}_{sin}^{i} \times \mathbf{n}_{cos}^{t(j)}] ) \sin\mathbf{k} \cdot \mathbf{r}_{ij} \\
 & + ( -J \mathbf{n}_{sin}^{i} \cdot \mathbf{n}_{sin}^{t(j)} + \mathbf{D}_{ij} \cdot [\mathbf{n}_{sin}^{i} \times \mathbf{n}_{sin}^{t(j)}] ) \cos\mathbf{k} \cdot \mathbf{r}_{ij} \} \cos\xi_{i}\cos\xi_{t(j)} \\
 & - N_{cell} g\mu_B \sum_{i=1}^{4} \mathbf{H} \cdot \mathbf{n}_{rot}^{i} \sin\xi_{i} .
\end{array}
\end{equation}
Using Eq. (\ref{eq:basis2}) we can rewrite the energy up to second order terms on $|\mathbf{k}| \sim (\omega_i - \omega_{t(j)}) \sim \gamma \sim (D/J)$:
\begin{equation}
\label{eq:EhelixH2}
\begin{array}{ll}
E_{helix,H} = & \frac{N_{cell}}{4} \sum_{i=1}^{4} \sum_{j=1}^{6} \{ -2J \cos(\xi_{i} - \xi_{t(j)}) \\
 & + 2 \left( \frac{J}{2} (\omega_i - \omega_{t(j)} - \mathbf{k} \cdot \mathbf{r}_{ij})^2 - (\mathbf{D}_{ij} \cdot \mathbf{n}_3) (\omega_i - \omega_{t(j)} - \mathbf{k} \cdot \mathbf{r}_{ij}) \right) \cos\xi_{i}\cos\xi_{t(j)} \\
 & + \left( \frac{J}{2} (\vec{\gamma}_i - \vec{\gamma}_{t(j)})^2 - \mathbf{D}_{ij} \cdot (\vec{\gamma}_i - \vec{\gamma}_{t(j)}) \right) (\cos\xi_{i} \cos\xi_{t(j)} + 2\sin\xi_{i} \sin\xi_{t(j)}) \} \\
 & - N_{cell} g\mu_B (\mathbf{H} \cdot \mathbf{n}_3) \sum_{i=1}^{4} \sin\xi_{i} .
\end{array}
\end{equation}
The last term has been obtained supposing that $g\mu_B H \sim D^2/J$, which allows us to use approximation $\mathbf{n}_{rot}^t \approx \mathbf{n}_3$. The first term in curly brackets has zero order on $(D/J)$ and achieves the minumum, when
\begin{equation}
\label{eq:xi}
\xi_1 = \xi_2 = \xi_3 = \xi_4 \equiv \xi .
\end{equation}
Then, we can rewrite the energy as
\begin{equation}
\label{eq:EhelixH3}
E_{helix,H} = N_{cell} ( -12J + \epsilon_1 \cos^2\xi + \epsilon_2 (1 + \sin^2\xi) - 4 g\mu_B (\mathbf{H} \cdot \mathbf{n}_3) \sin\xi ) ,
\end{equation}
where $\epsilon_1$ and $\epsilon_2$ are the same as in the previous section. The minimum of the energy (\ref{eq:EhelixH3}) is achieved, when $\mathbf{H} \parallel \mathbf{n}_3$ and $\mathbf{n}_3 = \mathbf{n}_\mathbf{k}$. Then,
\begin{equation}
\label{eq:EhelixH4}
E_{helix,H} = N_{cell} J \left( -12 - 2\frac{(D_x+D_z)^2}{J^2} - \frac{1}{3} \frac{(D_x-2D_y-D_z)^2}{J^2} \cos^2\xi - 4 \frac{g\mu_B H}{J} \sin\xi \right) ,
\end{equation}
or, after minimization on $\xi$,
\begin{equation}
\label{eq:EhelixH5}
E_{helix,H} = N_{cell} J \left( -12 - 2\frac{(D_x+D_z)^2}{J^2} - \frac{1}{3} \frac{(D_x-2D_y-D_z)^2}{J^2} - \frac{12 (g\mu_B H)^2}{(D_x-2D_y-D_z)^2} \right) ,
\end{equation}
and
\begin{equation}
\label{eq:sinxi}
\sin\xi = \frac{6 J g\mu_B H}{(D_x-2D_y-D_z)^2} .
\end{equation}

The helix vanishes, when $\xi = \frac{\pi}{2}$, which corresponds to magnetic field
\begin{equation}
\label{eq:H1} H_c = \frac{(D_x-2D_y-D_z)^2}{6 J
g\mu_B} .
\end{equation}
(this field, unwinding the helix, is usually
called $H_{c2}$ \cite{Grigoriev06}).

Thus, in the approximation used here, the wave
vector $\mathbf{k}$ is directed along the applied
magnetic field. The spin structure for each
sublattice obtains a permanent component along
its rotation axis $\mathbf{n}_{rot}$, which is
different from direction $\mathbf{n}_\mathbf{k}
\parallel \mathbf{H}$, see Fig.~\ref{fig1}(b).
The helices become conical, and as the magnetic
field increases up to $H_c$, the permanent
components grow proportionally to the field
value, Eq.~(\ref{eq:sinxi}), whereas the rotated
components decrease correspondingly. It should
be noted that if the field is less than $H_c$, it
does not affect the axes $\mathbf{n}_{rot}$ and
they have the same directions, defined by
Eq.~(\ref{eq:psi3}), as in the absence of
magnetic field.

\section{Coarse grain approximation}

In the phenomenological theory, the energy of
chiral magnetics is usually written as
\begin{equation}
\label{eq:Ephenomenological}
E = \int \left( {\cal J} \frac{\partial M_n}{\partial r_m} \frac{\partial M_n}{\partial r_m} + {\cal D} \mathbf{M} \cdot [\vec{\nabla} \times \mathbf{M}] \right) d\mathbf{r} ,
\end{equation}
where $\mathbf{M(r)}$ is the local magnetic
moment changing smoothly in space. Below it will
be shown how this equation and phenomenological
coefficients $\cal J$ and $\cal D$ arise from
microscopic consideration based on
Eq.~(\ref{eq:energy2}).

Let us suppose that the typical scale at which a considerable change of magnetic moment occures is much bigger than the size of the unit cell, and any two neighboring spins are rotated relative to each another by a small angle of order $(D/J)$. Let $\mathbf{s}_i$ and $\mathbf{s}_j$ be the neighboring spins and $\mathbf{s}_j = R(\vec{\zeta}_{ij}) \mathbf{s}_i$, or
\begin{equation}
\label{eq:sisj1}
\mathbf{s}_j =\mathbf{s}_i \cos\zeta_{ij} + \frac{\sin\zeta_{ij}}{\zeta_{ij}} [\vec{\zeta}_{ij} \times \mathbf{s}_i] + \frac{1-\cos\zeta_{ij}}{\zeta_{ij}^2} \vec{\zeta}_{ij} (\vec{\zeta}_{ij} \cdot \mathbf{s}_i) ,
\end{equation}
where $\zeta_{ij} \equiv |\vec{\zeta}_{ij}|$. Then,
\begin{equation}
\label{eq:sisj2}
\mathbf{s}_i \cdot \mathbf{s}_j \approx 1 - \zeta_{ij}^2 / 2 + (\mathbf{s}_i \cdot \vec{\zeta}_{ij})^2 / 2 = 1 - \zeta_{ij\perp}^2 / 2 ,
\end{equation}
\begin{equation}
\label{eq:sisj3}
[\mathbf{s}_i \times \mathbf{s}_j] \approx \vec{\zeta}_{ij} - \mathbf{s}_i (\mathbf{s}_i \cdot \vec{\zeta}_{ij}) = \vec{\zeta}_{ij\perp} ,
\end{equation}
where $\vec{\zeta}_{ij\perp}$ is the part of vector $\vec{\zeta}_{ij}$ perpendicular to unit vector $\mathbf{s}_i$.

If spins vary slowly along the crystal, we can take a domain, which is much bigger than a unit cell, but small enough as compared with the scale at wich a considerable change of spin occures. The spins within this domain are almost unidirectional. Consequently, all the vectors $\vec{\zeta}_{ij\perp}$ nearly lie in a plane. Let us neglect this ``nearly'' and suppose that all $\vec{\zeta}_{ij\perp}$ belong to the plane perpendicular to some mean spin $\mathbf{s}$. In addition, staying in the domain, we can discard the parts of $\vec{\zeta}_{ij}$ parallel to $\mathbf{s}$ and suppose that $\vec{\zeta}_{ij\perp} = \vec{\zeta}_{ij}$.

Another simplification can be achieved with taking into account that a spin does not change upon round a close chain of atoms. This means that the sum of small angles $\vec{\zeta}_{ij}$ corresponding to the bonds of the chain should be equal to zero. For instance, for the triangle formed by spins $\mathbf{s}_1$, $\mathbf{s}_2$ and $\mathbf{s}_3$ we obtain the following condition
\begin{equation}
\label{eq:3angle}
\vec{\zeta}_{12} + \vec{\zeta}_{23} + \vec{\zeta}_{31} = 0 ,
\end{equation}
which is evidently satisfied, if $\vec{\zeta}_{ij} = \vec{\zeta}_j - \vec{\zeta}_i$, where $\vec{\zeta}_i$ and $\vec{\zeta}_j$ are perpendicular to $\mathbf{s}$ vectors associated with $i$-th and $j$-th atoms.

With the approximations accepted, using the expressions (\ref{eq:sisj2}) and (\ref{eq:sisj3}), the energy (\ref{eq:energy2}) in the absence of magnetic field can be now rewritten as
\begin{equation}
\label{eq:energy3}
E = \frac{1}{2} \sum_i \sum_{j=1}^{6} ( -J + \frac{J}{2} (\vec{\zeta}_j - \vec{\zeta}_i)^2 + \mathbf{D}_{ij} \cdot (\vec{\zeta}_j - \vec{\zeta}_i) ) .
\end{equation}

The minimization of the energy on the vector $\vec{\zeta}_i$ associated with $i$-th atom and belonging to the plane perpendicular to $\mathbf{s}$ gives
\begin{equation}
\label{eq:zeta1}
\frac{\partial E}{\partial \vec{\zeta}_i} = - \sum_{j=1}^{6} ( J (\vec{\zeta}_j - \vec{\zeta}_i) + \mathbf{D}_{ij\perp} ) = 0 ,
\end{equation}
or
\begin{equation}
\label{eq:zeta2}
\vec{\zeta}_i = \frac{1}{6} \sum_{j=1}^{6} \vec{\zeta}_j + \frac{1}{6J} \left( \sum_{j=1}^{6} \mathbf{D}_{ij} \right)_{\perp} = \frac{1}{6} \sum_{j=1}^{6} \vec{\zeta}_j + \frac{1}{\sqrt{3}} \frac{D_x+D_z}{J} \mathbf{e}_{t(i)\perp} ,
\end{equation}
where symbol ``$\perp$'' designates the projection of a vector to the plane perpendicular to $\mathbf{s}$, $\mathbf{e}_{t(i)}$ is one of the vectors (\ref{eq:e}). The last therm in the right part of (\ref{eq:zeta2}) reflects the evident fact that the sum of DM vectors associated with all the bonds of an atom should be parallel to the 3-fold axis to which the atomic position belongs.

Let us introduce a new vector
$\tilde{\vec{\zeta}}_i$ connected with the old
one by equation
\begin{equation}
\label{eq:zeta3}
\vec{\zeta}_i  = \tilde{\vec{\zeta}}_i + A \mathbf{e}_{t(i)\perp} .
\end{equation}
where $A$ is a constant. Then,
\begin{equation}
\label{eq:zeta4}
\tilde{\vec{\zeta}}_i = \frac{1}{6} \sum_{j=1}^{6} \tilde{\vec{\zeta}}_j + \left( -\frac{4}{3} A + \frac{1}{\sqrt{3}} \frac{D_x+D_z}{J} \right) \mathbf{e}_{t(i)\perp} ,
\end{equation}
where it has been taken into account that an atom belonging to one of four sublattices has two neighbors at each of three remaining ones, so
\begin{equation}
\label{eq:sume}
\sum_{j=1}^{6} \mathbf{e}_{t(j)} = -2\mathbf{e}_{t(i)} .
\end{equation}
Choosing
\begin{equation}
\label{eq:A}
A = \frac{\sqrt{3}}{4} \frac{D_x+D_z}{J} ,
\end{equation}
we can now write
\begin{equation}
\label{eq:zeta5}
\vec{\zeta}_i = \tilde{\vec{\zeta}}_i + \frac{\sqrt{3}}{4} \frac{D_x+D_z}{J} \mathbf{e}_{t(i)\perp} ,
\end{equation}
\begin{equation}
\label{eq:zeta6}
\tilde{\vec{\zeta}}_i = \frac{1}{6} \sum_{j=1}^{6} \tilde{\vec{\zeta}}_j,
\end{equation}
and, after the substitution of Eq. (\ref{eq:zeta5}) into Eq. (\ref{eq:energy3}),
\begin{equation}
\label{eq:energy4}
\begin{array}{ll}
E = & \frac{1}{2} \sum_i \sum_{j=1}^{6} \{ -J + \frac{J}{2} (\tilde{\vec{\zeta}}_j - \tilde{\vec{\zeta}}_i)^2 + \tilde{\mathbf{D}}_{ij} \cdot (\tilde{\vec{\zeta}}_j - \tilde{\vec{\zeta}}_i) \} \\
 & + \frac{1}{2} \sum_i \sum_{j=1}^{6} \{ \frac{3}{32} \frac{(D_x+D_z)^2}{J} (\mathbf{e}_{t(j)} - \mathbf{e}_{t(i)})_{\perp}^2 + \frac{\sqrt{3}}{4} \frac{D_x+D_z}{J} \mathbf{D}_{ij} \cdot (\mathbf{e}_{t(j)} - \mathbf{e}_{t(i)})_{\perp} \} ,
\end{array}
\end{equation}
where
\begin{equation}
\label{eq:Dnew1}
\begin{array}{ll}
\tilde{\mathbf{D}}_{ij} & = \mathbf{D}_{ij} + \frac{\sqrt{3}}{4} (D_x+D_z) (\mathbf{e}_{t(j)} - \mathbf{e}_{t(i)}) \\
 & = \mathbf{D}_{ij} + \frac{1}{8} \left( \sum_{j^{\prime}=1}^6 \mathbf{D}_{t(j)j^{\prime}} - \sum_{j^{\prime\prime}=1}^6 \mathbf{D}_{t(i)j^{\prime\prime}} \right).
\end{array}
\end{equation}
The first part of Eq. (\ref{eq:energy4}) is analogous to the expression (\ref{eq:energy3}), but with modificated DM vectors. It is important here that the combination $(\mathbf{e}_{t(j)} - \mathbf{e}_{t(i)})$ varies from bond to bond with the use of the same symmetry transformations of the point group $m\bar{3}$ as vector $\mathbf{D}_{ij}$ and, consequently, $\tilde{\mathbf{D}}_{ij}$. The vectors $\tilde{\mathbf{D}}_{ij}$ are characterized by condition $\tilde{D}_x + \tilde{D}_z = 0$, so they have only two independent components. For example, the modificated DM vector of the bond $(-2x, \frac{1}{2}, \frac{1}{2}-2x)$ is
\begin{equation}
\label{eq:Dnew2}
\tilde{\mathbf{D}} = (D_x, D_y, D_z) + \frac{\sqrt{3}}{4} (D_x+D_z) (\mathbf{e}_2 - \mathbf{e}_1) = \left( \frac{D_x-D_z}{2}, D_y, -\frac{D_x-D_z}{2} \right) .
\end{equation}

The second part of Eq. (\ref{eq:energy4}) contains the terms like
\begin{equation}
\label{eq:ab1}
\mathbf{a}_{\perp} \cdot \mathbf{b}_{\perp} = \mathbf{a} \cdot \mathbf{b} - (\mathbf{a} \cdot \mathbf{s}) (\mathbf{b} \cdot \mathbf{s}) ,
\end{equation}
depending on the direction of $\mathbf{s}$. The vectors $\mathbf{a}$ and $\mathbf{b}$ vary from bond to bond with the use of the symmetry transformations of the point group $23$, and the averaging over 12 bonds in the unit cell, as well as for the case of the averaging over the sphere, gives
\begin{equation}
\label{eq:ab2}
\left\langle \mathbf{a}_{\perp} \cdot \mathbf{b}_{\perp} \right\rangle = \mathbf{a} \cdot \mathbf{b} - \frac{1}{12} \sum_{n=1}^{12} ( (R_n \mathbf{a}) \cdot \mathbf{s} ) ( (R_n \mathbf{b}) \cdot \mathbf{s} ) = \mathbf{a} \cdot \mathbf{b} - \frac{1}{3} \mathbf{a} \cdot \mathbf{b} = \frac{2}{3} \mathbf{a} \cdot \mathbf{b} ,
\end{equation}
with $R_n$ being the rotation matrices of the group $23$. Thereby, the mean contribution of the second part of Eq. (\ref{eq:energy4}) to the energy of a bond is
\begin{equation}
\label{eq:Ediv}
\frac{2}{3} \{ \frac{3}{32} \frac{(D_x+D_z)^2}{J} (\mathbf{e}_2 - \mathbf{e}_1)^2 + \frac{\sqrt{3}}{4} \frac{D_x+D_z}{J} (D_x, D_y, D_z) \cdot (\mathbf{e}_2 - \mathbf{e}_1) \} = -\frac{(D_x+D_z)^2}{6J} ,
\end{equation}
and we can now rewrite the expression (\ref{eq:energy4}) for the energy as
\begin{equation}
\label{eq:energy5}
E = \frac{1}{2} \sum_i \sum_{j=1}^{6} \{ \frac{J}{2} (\tilde{\vec{\zeta}}_j - \tilde{\vec{\zeta}}_i)^2 + \tilde{\mathbf{D}}_{ij} \cdot (\tilde{\vec{\zeta}}_j - \tilde{\vec{\zeta}}_i) \} + N_{cell} J \left( -12 - 2\left( \frac{D_x+D_z}{J} \right)^2 \right) .
\end{equation}

Let us introduce now smooth continuous vector fields $\hat{\vec{\zeta}}_1 (\mathbf{r})$, $\hat{\vec{\zeta}}_2 (\mathbf{r})$, $\hat{\vec{\zeta}}_3 (\mathbf{r})$ and $\hat{\vec{\zeta}}_4 (\mathbf{r})$, connecting with $\tilde{\vec{\zeta}}_i$ by the equation
\begin{equation}
\label{eq:hat1}
\tilde{\vec{\zeta}}_i = \hat{\vec{\zeta}}_{t(i)} (\mathbf{r}_i) .
\end{equation}
We can suppose that functions $\hat{\vec{\zeta}}_t$ change slowly enough with coordinates and postulate the homogeneity condition
\begin{equation}
\label{eq:homogeneity}
\frac{\partial (\hat{\vec{\zeta}}_1)_n}{\partial r_m} = \frac{\partial (\hat{\vec{\zeta}}_2)_n}{\partial r_m} = \frac{\partial (\hat{\vec{\zeta}}_3)_n}{\partial r_m} = \frac{\partial (\hat{\vec{\zeta}}_4)_n}{\partial r_m} \equiv \frac{\partial \hat{\zeta}_n}{\partial r_m} , \phantom{x} m, n = 1, 2, 3,
\end{equation}
reflecting the fact that the divergence between sublattices is preserved over big distances. Then, we can write Eq. (\ref{eq:zeta6}) for 1st atom in a cell in the explicit form as
\begin{equation}
\label{eq:hat2}
\begin{array}{lll}
\hat{\vec{\zeta}}_1 (x, x, x) & = & \frac{1}{6} \left( \hat{\vec{\zeta}}_2 (-x, \frac{1}{2}+x, \frac{1}{2}-x) + \hat{\vec{\zeta}}_2 (-x, -\frac{1}{2}+x, \frac{1}{2}-x) \right. \\
 & & + \hat{\vec{\zeta}}_3 (\frac{1}{2}-x, -x, \frac{1}{2}+x) + \hat{\vec{\zeta}}_3 (\frac{1}{2}-x, -x, -\frac{1}{2}+x) \\
 & & \left. + \hat{\vec{\zeta}}_4 (\frac{1}{2}+x, \frac{1}{2}-x, -x) + \hat{\vec{\zeta}}_4 (-\frac{1}{2}+x, \frac{1}{2}-x, -x) \right) \\
 & = & \frac{1}{3} \left( \hat{\vec{\zeta}}_2 (x, x, x) + \hat{\vec{\zeta}}_3 (x, x, x) + \hat{\vec{\zeta}}_4 (x, x, x) \right) + \frac{1}{6} \left( \sum_{j=1}^6 \mathbf{r}_{1j} \cdot \vec{\nabla} \right) \hat{\vec{\zeta}} (x, x, x) .
\end{array}
\end{equation}
The last equation is correct not only for $\mathbf{r} = (x, x, x)$ but also for arbitrary $\mathbf{r}$ due to homogeneity condition. So, we can write Eq. (\ref{eq:hat2}) and analogous equations for other atoms in a cell, supposing that all $\hat{\vec{\zeta}}_i$ are taken at the same point:
\begin{equation}
\label{eq:hat3}
\left\{
\begin{array}{ll}
\hat{\vec{\zeta}}_1  & = \frac{1}{3} \left( \hat{\vec{\zeta}}_2 + \hat{\vec{\zeta}}_3 + \hat{\vec{\zeta}}_4 \right) + \frac{1}{6} \left( \sum_{j=1}^6 \mathbf{r}_{1j} \cdot \vec{\nabla} \right) \hat{\vec{\zeta}} , \\
\hat{\vec{\zeta}}_2  & = \frac{1}{3} \left( \hat{\vec{\zeta}}_1 + \hat{\vec{\zeta}}_3 + \hat{\vec{\zeta}}_4 \right) + \frac{1}{6} \left( \sum_{j=1}^6 \mathbf{r}_{2j} \cdot \vec{\nabla} \right) \hat{\vec{\zeta}} , \\
\hat{\vec{\zeta}}_3  & = \frac{1}{3} \left( \hat{\vec{\zeta}}_1 + \hat{\vec{\zeta}}_2 + \hat{\vec{\zeta}}_4 \right) + \frac{1}{6} \left( \sum_{j=1}^6 \mathbf{r}_{3j} \cdot \vec{\nabla} \right) \hat{\vec{\zeta}} , \\
\hat{\vec{\zeta}}_4  & = \frac{1}{3} \left( \hat{\vec{\zeta}}_1 + \hat{\vec{\zeta}}_2 + \hat{\vec{\zeta}}_3 \right) + \frac{1}{6} \left( \sum_{j=1}^6 \mathbf{r}_{4j} \cdot \vec{\nabla} \right) \hat{\vec{\zeta}} .
\end{array}
\right.
\end{equation}
The system (\ref{eq:hat3}) is degenerate on variables $\hat{\vec{\zeta}}_1$, $\hat{\vec{\zeta}}_2$, $\hat{\vec{\zeta}}_3$, $\hat{\vec{\zeta}}_4$ and satisfied, when
\begin{equation}
\label{eq:hat4}
\hat{\vec{\zeta}}_t - \hat{\vec{\zeta}}_{t^\prime} = \frac{1}{8} \left( \sum_{j=1}^6 \mathbf{r}_{tj} \cdot \vec{\nabla} - \sum_{j^{\prime}=1}^6 \mathbf{r}_{t^{\prime}j^{\prime}} \cdot \vec{\nabla} \right) \hat{\vec{\zeta}} .
\end{equation}
Let us allocate two parts from expression (\ref{eq:energy5}) for the energy associated with a unit cell,
\begin{equation}
\label{eq:Eexch1}
E_{cell,exch} = \frac{J}{4} \sum_{i=1}^{4} \sum_{j=1}^{6} (\hat{\vec{\zeta}}_{t(j)} (\mathbf{r}_j) - \hat{\vec{\zeta}}_{i} (\mathbf{r}_i))^2 = \frac{J}{4} \sum_{i=1}^{4} \sum_{j=1}^{6} (\hat{\vec{\zeta}}_{t(j)} - \hat{\vec{\zeta}}_{i} + (\mathbf{r}_{ij} \cdot \vec{\nabla}) \hat{\vec{\zeta}})^2
\end{equation}
and
\begin{equation}
\label{eq:EDM1}
E_{cell,DM} = \frac{1}{2} \sum_{i=1}^{4} \sum_{j=1}^{6} \tilde{\mathbf{D}}_{ij} \cdot (\hat{\vec{\zeta}}_{t(j)} (\mathbf{r}_j) - \hat{\vec{\zeta}}_{i} (\mathbf{r}_i)) = \frac{1}{2} \sum_{i=1}^{4} \sum_{j=1}^{6} \tilde{\mathbf{D}}_{ij} \cdot (\hat{\vec{\zeta}}_{t(j)} - \hat{\vec{\zeta}}_{i} + (\mathbf{r}_{ij} \cdot \vec{\nabla}) \hat{\vec{\zeta}}) .
\end{equation}
With the use of Eq. (\ref{eq:hat4}) we can rewrite these parts as
\begin{equation}
\label{eq:Eexch2}
E_{cell,exch} = - \frac{J}{16} \sum_{i=1}^{4} \left( \left( \sum_{j=1}^{6} \mathbf{r}_{ij} \cdot \vec{\nabla} \right) \hat{\vec{\zeta}} \right)^2 + \frac{J}{4} \sum_{i=1}^{4} \sum_{j=1}^{6} \left( (\mathbf{r}_{ij} \cdot \vec{\nabla}) \hat{\vec{\zeta}} \right)^2
\end{equation}
and
\begin{equation}
\label{eq:EDM2}
E_{cell,DM} = \frac{1}{2} \sum_{i=1}^{4} \sum_{j=1}^{6}  (\mathbf{r}_{ij} \cdot \vec{\nabla}) (\tilde{\mathbf{D}}_{ij} \cdot \hat{\vec{\zeta}}) - \frac{1}{8} \sum_{i=1}^{4} \left( \sum_{j=1}^{6} \mathbf{r}_{ij} \cdot \vec{\nabla} \right) \left( \sum_{j^{\prime}=1}^{6} \tilde{\mathbf{D}}_{ij^{\prime}} \cdot \hat{\vec{\zeta}} \right) .
\end{equation}
The second part of (\ref{eq:EDM2}) vanishes, because
\begin{equation}
\label{eq:sumD}
\sum_{j=1}^{6} \tilde{\mathbf{D}}_{ij} = 0 .
\end{equation}

Using
\begin{equation}
\label{eq:sumr}
\sum_{j=1}^{6} \mathbf{r}_{ij} = \sqrt{3} (1 - 8x) \mathbf{e}_i ,
\end{equation}
the first part of (\ref{eq:Eexch2}) can be calculated,
\begin{equation}
\label{eq:EexchPart1}
- \frac{J}{16} \sum_{i=1}^{4} \left( \left( \sum_{j=1}^{6} \mathbf{r}_{ij} \cdot \vec{\nabla} \right) \hat{\vec{\zeta}} \right)^2 = - \frac{J}{4} (1 - 8x)^2 \frac{\partial \hat{\zeta}_n}{\partial r_m} \frac{\partial \hat{\zeta}_n}{\partial r_m} .
\end{equation}
The second part of (\ref{eq:Eexch2}) can be expressed as
\begin{equation}
\label{eq:EexchPart2}
\frac{J}{4} \sum_{\alpha=1}^{24} R_{kl}^{\alpha} R_{mn}^{\alpha} ((\mathbf{r}_{12})_l \nabla_k \hat{\vec{\zeta}}) \cdot ((\mathbf{r}_{12})_n \nabla_m \hat{\vec{\zeta}}) = J (1 - 4x + 16x^2) \frac{\partial \hat{\zeta}_n}{\partial r_m} \frac{\partial \hat{\zeta}_n}{\partial r_m} ,
\end{equation}
where summation is taken over all elements $R^{\alpha}$ of the point symmetry group $m\bar{3}$, and we have used that
\begin{equation}
\label{eq:sumRR}
\sum_{\alpha=1}^{24} R_{kl}^{\alpha} R_{mn}^{\alpha} = 8 \delta_{km} \delta_{ln}
\end{equation}
and
\begin{equation}
\label{eq:rsquare}
(\mathbf{r}_{12})^2 = \frac{1}{2} (1 - 4x + 16x^2) .
\end{equation}
So, finally,
\begin{equation}
\label{eq:Eexch3}
E_{cell,exch} = \frac{3J}{4} \frac{\partial \hat{\zeta}_n}{\partial r_m} \frac{\partial \hat{\zeta}_n}{\partial r_m} .
\end{equation}
Similarly, Eq. (\ref{eq:EDM2}) can be rewritten as
\begin{equation}
\label{eq:EDM3a}
E_{cell,DM} = \sum_{\alpha=1}^{24} R_{kl}^{\alpha} R_{mn}^{\alpha} (\mathbf{r}_{12})_l (\tilde{\mathbf{D}}_{12})_n \nabla_k \hat{\zeta}_m = 4 (\mathbf{r}_{12} \cdot \tilde{\mathbf{D}}_{12}) \frac{\partial \hat{\zeta}_n}{\partial r_n}
\end{equation}
or
\begin{equation}
\label{eq:EDM3b}
E_{cell,DM} = - (D_x-2D_y-D_z) \frac{\partial \hat{\zeta}_n}{\partial r_n} .
\end{equation}

It should be noted that, whereas Eqs.
(\ref{eq:Eexch2}) and (\ref{eq:EDM2}) contain
parameter $x$, the resulting expressions
(\ref{eq:Eexch3}) and (\ref{eq:EDM3b}) do not
depend on $x$. This looks like a coincidence, but
it is not so. Indeed, the initial statement of
the problem do not include atomic coordinates
(e.g., see Eq. (\ref{eq:energy1})), and therefore
the coordinates will not appear in results (at
least until strong changes of the coordinates
changing the topology of the net of nearest
neighbors).

The substitution of Eqs. (\ref{eq:Eexch3}) and (\ref{eq:EDM3b}) into Eq. (\ref{eq:energy5}) gives
\begin{equation}
\label{eq:energy6}
E = \int \left( {\cal J} \frac{\partial \hat{\zeta}_n}{\partial r_m} \frac{\partial \hat{\zeta}_n}{\partial r_m} - {\cal D} \frac{\partial\hat{\zeta}_n}{\partial r_n} \right) d\mathbf{r} + N_{cell} J \left( -12 - 2\left( \frac{D_x+D_z}{J} \right)^2 \right) ,
\end{equation}
with ${\cal J}$ and ${\cal D}$ given by the following equations
\begin{equation}
\label{eq:J1}
{\cal J} = \frac{3J}{4} ,
\end{equation}
\begin{equation}
\label{eq:D1}
{\cal D} = - 4 (\mathbf{r}_{12} \cdot \tilde{\mathbf{D}}_{12}) = D_x-2D_y-D_z .
\end{equation}
The summation over all unit cells of the crystal has been replaced with the integration over the crystal volume.

Let us suppose that the vector field $\hat{\vec{\zeta}}$ is connected with a spin field $\hat{\mathbf{s}}$:
\begin{equation}
\label{eq:szeta1}
\hat{\mathbf{s}} \approx \mathbf{s}_0 + [\hat{\vec{\zeta}} \times \mathbf{s}_0] ,
\end{equation}
where $\mathbf{s}_0$ is a constant vector, $\hat{\vec{\zeta}} \perp \mathbf{s}_0$. Then,
\begin{equation}
\label{eq:szeta2}
\frac{\partial \hat{s}_n}{\partial r_m} \frac{\partial \hat{s}_n}{\partial r_m} \approx \frac{\partial \hat{\zeta}_n}{\partial r_m} \frac{\partial \hat{\zeta}_n}{\partial r_m}
\end{equation}
and
\begin{equation}
\label{eq:szeta3}
\hat{\mathbf{s}} \cdot [\vec{\nabla} \times \hat{\mathbf{s}}] \approx - \frac{\partial \hat{\zeta}_n}{\partial x_n} .
\end{equation}
So, we can rewrite Eq. (\ref{eq:energy6}) for the energy as
\begin{equation}
\label{eq:energy7}
E = \int \left( {\cal J} \frac{\partial \hat{s}_n}{\partial r_m} \frac{\partial \hat{s}_n}{\partial r_m} + {\cal D} \hat{\mathbf{s}} \cdot [\vec{\nabla} \times \hat{\mathbf{s}}] \right) d\mathbf{r} + N_{cell} J \left( -12 - 2\left( \frac{D_x+D_z}{J} \right)^2 \right) ,
\end{equation}
the first part of which coincides with the usually used energy in the phenomenological theory with the elastic constants ${\cal J}$ and ${\cal D}$ related to the the constants of microscopic interactions by Eqs. (\ref{eq:J1}) and (\ref{eq:D1}).

The wave number for the helix structure can be now found using the phenomenological approach,
\begin{equation}
\label{eq:k3}
k = \frac{{\cal D}}{2{\cal J}} = \frac{2}{3} \left( \frac{D_x-2D_y-D_z}{J} \right) ,
\end{equation}
which coincides with Eq. (\ref{eq:k2}).

Thus, we have passed from the discrete spin distribution of the microscopic theory to the continuous spin field $\hat{\mathbf{s}}$ of the coarse grain theory. In order to complete our reasoning we can also consider the way back from the continuous approximation to dicrete spin distibution.

Let $\hat{\mathbf{s}}(\mathbf{r})$ be a continuous spin field from the phenomenological theory. Within a small domain of the crystal we can associate with $\hat{\mathbf{s}}$ a vector field $\hat{\vec{\zeta}}$ using the transformation inverse to (\ref{eq:szeta1}):
\begin{equation}
\label{eq:szeta4}
\hat{\vec{\zeta}} = [\mathbf{s}_0 \times \hat{\mathbf{s}}] .
\end{equation}
Here $\mathbf{s}_0$ is some mean value of $\hat{\mathbf{s}}$, which is constant over the domain. Then,
\begin{equation}
\label{eq:szeta5}
\frac{\partial \hat{\vec{\zeta}}}{\partial r_n} = [\mathbf{s}_0 \times \frac{\partial \hat{\mathbf{s}}}{\partial r_n}] \approx [\hat{\mathbf{s}} \times \frac{\partial \hat{\mathbf{s}}}{\partial r_n}] .
\end{equation}

Let us define continuous fields $\hat{\vec{\zeta}}_1$, $\hat{\vec{\zeta}}_2$, $\hat{\vec{\zeta}}_3$ and $\hat{\vec{\zeta}}_4$ for four sublattices of magnetic atoms by formula
\begin{equation}
\label{eq:hat5}
\hat{\vec{\zeta}}_t = \hat{\vec{\zeta}} + \frac{1}{8} \sum_{j=1}^6 (\mathbf{r}_{tj} \cdot \vec{\nabla}) \hat{\vec{\zeta}} ,
\end{equation}
obviously obeying Eq. (\ref{eq:hat4}). Thus $\hat{\vec{\zeta}}$ is the average of these four fields,
\begin{equation}
\label{eq:hat6}
\hat{\vec{\zeta}} = \frac{1}{4} (\hat{\vec{\zeta}}_1 + \hat{\vec{\zeta}}_2 + \hat{\vec{\zeta}}_3 + \hat{\vec{\zeta}}_4) .
\end{equation}
Then, using Eqs. (\ref{eq:hat1}) and (\ref{eq:zeta5}) we can express vector $\vec{\zeta}_i$ for individual atom trough $\hat{\vec{\zeta}}$,
\begin{equation}
\label{eq:zeta7}
\vec{\zeta}_i = \hat{\vec{\zeta}} (\mathbf{r}_i) + \frac{\sqrt{3}}{4} \frac{D_x+D_z}{J} \mathbf{e}_{t(i)\perp} (\mathbf{r}_i) + \frac{\sqrt{3}}{8} (1 - 8x) (\mathbf{e}_{t(i)} \cdot \vec{\nabla}) \hat{\vec{\zeta}} (\mathbf{r}_i) ,
\end{equation}
where
\begin{equation}
\label{eq:e2}
\mathbf{e}_{t(i)\perp} (\mathbf{r}_i) = \mathbf{e}_{t(i)} - \hat{\mathbf{s}} (\mathbf{r}_i) (\hat{\mathbf{s}}  (\mathbf{r}_i) \cdot \mathbf{e}_{t(i)}) .
\end{equation}

The substitution of Eq. (\ref{eq:zeta7}) into the equation
\begin{equation}
\label{eq:szeta6}
\mathbf{s}_i = \hat{\mathbf{s}} + [\vec{\zeta}_i \times \hat{\mathbf{s}}]
\end{equation}
gives us the following expression for the spin of individual magnetic atom
\begin{equation}
\label{eq:sdiscrete}
\mathbf{s}_i = \hat{\mathbf{s}} (\mathbf{r}_i) + \frac{\sqrt{3}}{4} \frac{D_x+D_z}{J} [\mathbf{e}_{t(i)} \times \hat{\mathbf{s}} (\mathbf{r}_i)] + \frac{\sqrt{3}}{8} (1 - 8x) (\mathbf{e}_{t(i)} \cdot \vec{\nabla}) \hat{\mathbf{s}} (\mathbf{r}_i) .
\end{equation}
This equation is valid for any smooth
continuous field
$\hat{\mathbf{s}}(\mathbf{r})$. For
one-dimensional helices, the first term in
Eq.~(\ref{eq:sdiscrete}), which is common for all
sublattices, describes spin rotation in the
plane perpendicular to the wave vector
$\mathbf{k}$, the second one defines a spin
component parallel to $\mathbf{k}$, and second
and third ones taken together do a shift of the
phase of the rotation for each sublattice. For
instance, if one chooses $\hat{\mathbf{s}}$ in
the form of
\begin{equation}
\label{eq:shat}
\hat{\mathbf{s}}(\mathbf{r}) = \mathbf{n}_1 \cos\mathbf{k} \cdot \mathbf{r} + \mathbf{n}_2 \sin\mathbf{k} \cdot \mathbf{r} ,
\end{equation}
then $\mathbf{s}_i = \mathbf{s}^{t(i)}(\mathbf{r}_i)$ has the form (\ref{eq:s3}).

\section{Discussion and summary}
\label{discussion}

As it is shown above, in the approximation linear
on $(D/J)$, only two components of DM vectors are
significant for the considered magnetic structure
of MnSi: $D_x-2D_y-D_z$ and $D_x+D_z$. The third
component, $D_x+D_y-D_z$, appears in upper
approximations. The pseudo-scalar constant ${\cal
D}$ used in the phenomenological theory is found
to be proportional to the first component,
$D_x-2D_y-D_z$. In fact, we could expect that
only this combination of $\mathbf{D}$-vector
coordinates played a role in the simplest
microscopic approximation. Surprisingly, it has
been found that another component, $D_x+D_z$, is
also very important both because of its
contribution into magnetic energy and its
influence on spins orientations. This effect is
very similar to ``frozen'' optical phonon
displacements of atoms in crystals. Whereas the
former component is responsible for the long
scale chiral magnetic structure, the latter leads
to a canting of spins within a unit cell and this
canting could be in principle measured
experimentally using diffraction methods
\cite{Dmitrienko11}. For instance, for the
magnetic fields stronger than $H_c$, the residual
splay of spins results in non-zero value of the
neutron scattering amplitude for reflections of
$00\ell, \ell=2n+1$ type, forbidden both for
nuclear scattering and for ferromagnetic
scattering owing to $2_1$ screw axes.

An interesting result about the symmetry of the
problem can be revealed considering the inverse
structure, where magnetic atoms occupy positions
$(-x, -x, -x)$ rather than $(x, x, x)$. The
crystal thus obtained has the same space group,
$P2_13$, but both the crystals are related as
right- and left-hand structures, via the
inversion transformation. It can be easily seen
from the expression (\ref{eq:energy1}) that
energy does not change over this transformation,
if the spins in corresponding positions are equal
and so do Dzyaloshinskii--Moriya vectors
associated with corresponding bonds. This means
that DM vectors are really pseudovectors. Also it
is well understood \cite{Grigoriev09,Grigoriev10}
that the inversion should change the chirality of
the structure and, consequently, the sign of
phenomenological constant ${\cal D}$ and wave
number $k$. However it is not so evident from
Eq.~(\ref{eq:k3}) that $k$ changes its sign over
inversion, because $\mathbf{D}$ is a pseudovector
rather than a vector. We should rewrite the wave
number in an invariant form, including an
explicit dependence on the bond direction, in
order to make transparent the change of chirality
upon inversion:
\begin{equation}
\label{eq:k4}
k = - \frac{8(\mathbf{r}_{12} \cdot \tilde{\mathbf{D}}_{12})}{3J} .
\end{equation}
The pseudovector $\tilde{\mathbf{D}}_{12}$ does
not change its sign under inversion but
$\mathbf{r}_{12}$ does. Notice that in this form
the sign of $k$ does not depend on our choice
which of atoms is 1 or 2.

It is interesting to compare our results with
those ones obtained by Hopkinson and Kee
\cite{Hopkinson} using mean-field and real-space
rigid rotor minimization methods. According to
this work, for both methods, a sizable spin
helicity can be obtained only when the DM vectors
lie {\it parallel} to the corresponding Mn-Mn
bonds. Correspondingly, three components of DM
vector are considered: the first one, parallel to
the bond, the second one, perpendicular to the
triangle of bonds, to which this bond is
belonging, and the third one, lying in the plane
of the triangle, but perpendicular to the bond.
For instance, for the bond $\mathbf{r}_{12} =
(-2x, \frac{1}{2}, \frac{1}{2} - 2x)$ these three
directions can be written in explicit form as
$(-2x, \frac{1}{2}, \frac{1}{2} - 2x)$, $(1, 1,
-1)$ and $(1 - 2x, 4x - \frac{1}{2}, \frac{1}{2}
+ 2x)$, correspondingly, or, using $x = 0.138$,
as $(-0.276, 0.5, 0.224)$, $(1, 1, -1)$ and
$(0.724, 0.052, 0.776)$. The first component is
responsible for the helix structure, and
corresponding direction, $(-0.276, 0.5, 0.224)$,
is very close to direction $(-1, 2, 1)$ for
$D_x-2D_y-D_z$ component found in our work (the
angle between them is about 3.4 degrees). The
second component, which does not cant spins
\cite{Hopkinson}, coincides with our ``neutral''
component $D_x+D_y-D_z$ directed along the
corresponding 3-fold axis. The third component
leads to spin canting and rather small spin
spiraling, and its direction, $(0.724, 0.052,
0.776)$, is very close to direction $(1,0, 1)$
for $D_x+D_z$ component. Obviously, those
numerical results almost coincide with our
analytical solutions. An advantage of the
analytical solution is that it does not rely on
the exact bond directions and atomic coordinates;
that is natural from the theoretical point of
view because the atomic parameter $x$ does not
figure in Eq. \ref{eq:energy1}.

Moreover, different model considerations predict
that the superexchange DM vectors are either
perpendicular \cite{Keffer} or almost
perpendicular
\cite{Shekhtman93,Mazurenko05,Katsnelson10} to
the corresponding bonds. If this is also the case
in MnSi, we could expect pronounced frustration
of the helices. In any case, it is obvious that
modern {\it ab initio} calculations of the
Dyzaloshinskii-Moriya vectors would be very
important for MnSi-type crystals.

In conclusion, it is found which component of the
Dyzaloshinskii-Moriya vector determines the
long-range spiral ordering. The most interesting
point is that the local structure of helices can
be rather complicated owing to symmetry-driven
frustrations. We hope that qualitatively all the
considered features will appear also in more
sophisticated models taking into account the
itinerant nature of magnetism and thermal and
quantum properties of spiral ordering in MnSi.

\section*{Acknowledgements}
We are grateful to S.~V. Grigoriev and S.~V.
Maleyev for useful discussions. This work was
supported by two basic research programs of the
Presidium of the Russian Academy of Sciences:
``Thermophysics and Mechanics of Extreme Energy
Actions and the Physics of a Strongly Compressed
Substance'' and ``Resonant X-ray Diffraction and
Topography in Forbidden Reflections''.

\newpage
\section*{Tables}

\begin{table}[h]
\begin{center}
\begin{tabular}{|r|c|c|c|c|}
\hline
$n$ & $\mathbf{r}_{ij}$ & $t(i)$ & $t(j)$ & $\mathbf{D}_{ij}$ \\
\hline
$1$ & $(-2x, \frac{1}{2}, \frac{1}{2}-2x)$ & $1$ & $2$ & $(D_x, D_y, D_z)$ \\
\hline
$2$ & $(\frac{1}{2}-2x, -2x, \frac{1}{2})$ & $1$ & $3$ & $(D_z, D_x, D_y)$ \\
\hline
$3$ & $(\frac{1}{2}, \frac{1}{2}-2x, -2x)$ & $1$ & $4$ & $(D_y, D_z, D_x)$ \\
\hline
$4$ & $(2x, \frac{1}{2}, -\frac{1}{2}+2x)$ & $2$ & $1$ & $(-D_x, D_y, -D_z)$ \\
\hline
$5$ & $(-\frac{1}{2}, \frac{1}{2}-2x, 2x)$ & $2$ & $3$ & $(-D_y, D_z, -D_x)$ \\
\hline
$6$ & $(-\frac{1}{2}+2x, -2x, -\frac{1}{2})$ & $2$ & $4$ & $(-D_z, D_x, -D_y)$ \\
\hline
$7$ & $(-\frac{1}{2}+2x, 2x, \frac{1}{2})$ & $3$ & $1$ & $(-D_z, -D_x, D_y)$ \\
\hline
$8$ & $(-\frac{1}{2}, -\frac{1}{2}+2x, -2x)$ & $3$ & $2$ & $(-D_y, -D_z, D_x)$ \\
\hline
$9$ & $(2x, -\frac{1}{2}, \frac{1}{2}-2x)$ & $3$ & $4$ & $(-D_x, -D_y, D_z)$ \\
\hline
$10$ & $(\frac{1}{2}, -\frac{1}{2}+2x, 2x)$ & $4$ & $1$ & $(D_y, -D_z, -D_x)$ \\
\hline
$11$ & $(\frac{1}{2}-2x, 2x, -\frac{1}{2})$ & $4$ & $2$ & $(D_z, -D_x, -D_y)$ \\
\hline
$12$ & $(-2x, -\frac{1}{2}, -\frac{1}{2}+2x)$ & $4$ & $3$ & $(D_x, -D_y, -D_z)$ \\
\hline
\end{tabular}
\caption{\label{tableD} The bonds between
neighboring Mn atoms in MnSi crystal and
corresponding Dzyaloshinskii--Moriya vectors.
The listed 12 bonds are connected by symmetry
transformations of the point group $23$. Other 12
bonds appear owing to permutation of $i$-th and
$j$-th atoms ($\mathbf{r}_{ji} =
-\mathbf{r}_{ij}$, $\mathbf{D}_{ji} =
-\mathbf{D}_{ij}$).}
\end{center}
\end{table}

\newpage
\section*{Figures}

\newpage

\end{document}